\documentclass[a4paper,11pt]{article}

\usepackage{jcappub} 

\usepackage{placeins}
\usepackage{multirow}
\usepackage{booktabs}
\usepackage{xcolor}
\usepackage{lineno}
\usepackage{hyperref}

\title{\boldmath Neutron production induced by \texorpdfstring{$\alpha$}{alpha}-decay with Geant4}

\author[1]{E.~Mendoza,\note{Corresponding author.}}
\author{D.~Cano-Ott,}
\author{P.~Romojaro,}
\author{V.~Alcayne,}
\author{P.~Garc\'{i}a Abia,}
\author{V.~Pesudo,}
\author{L.~Romero}
\author{and R.~Santorelli}

\affiliation{CIEMAT, Avda. Complutense 40, 28040 Madrid, Spain}

\emailAdd{emilio.mendoza@ciemat.es}
\emailAdd{daniel.cano@ciemat.es}
\emailAdd{Pablo.Romojaro@ciemat.es}
\emailAdd{victor.alcayne@ciemat.es}
\emailAdd{pablo.garcia@ciemat.es}
\emailAdd{vicente.pesudo@ciemat.es}
\emailAdd{luciano.romero@ciemat.es}
\emailAdd{roberto.santorelli@ciemat.es}

\abstract{
Neutron production induced by ($\alpha$,n) reactions is important in various scenarios. One of the most relevant ones is related to underground dark matter experiments, where the neutrons produced by the $\alpha$-decay from radioactive contaminants in the detector materials can generate an irreducible background, limiting the sensitivity of the experiment. A precise estimate of the background due to these neutrons is crucial for the experiments currently taking data and for the design of the next generation detectors. In this work, we prove that Geant4 can be used to calculate neutron yields and energy spectra induced by $\alpha$-decay. These neutrons are produced according to the information compiled in data libraries originally written in the ENDF-6 format. In this article we also review the different databases available, showing the differences and similarities among them. Finally, we compare several Geant4 neutron production yields and spectra with experimental data and other codes.
}


\begin{document}
\maketitle 
\flushbottom

\section{Introduction \label{Sec:intro}}

Calculations of the neutron production induced by $\alpha$-decay through ($\alpha$,n) reactions are required in various fields. One such field is that of nuclear power technologies, in which the production of neutrons by actinide $\alpha$-decay plays an important role in various scenarios, such as the treatment of fresh and spent fuel~\cite{RED-IMPACT}, nuclear safeguards~\cite{NuclearSafeguards} or radiological protection~\cite{RadioProtection}. Another relevant example is given by the rare-events search experiments carried out in underground laboratories, which need to reach extremely low background conditions, minimizing signals from cosmic rays and natural radioactivity. Materials with very low radioactive contaminants (ppb or less) are typically used for the detector construction. However, some radioactivity is inevitably always present, which leads to the production of neutrons, among other particles. These neutrons can be produced by spontaneous fission, $\beta$n-decay, or by ($\alpha$,n) reactions initiated by $\alpha$ particles from the decay of nuclei belonging to the natural uranium and thorium decay chains. These reactions are especially important in underground dark matter experiments aimed at detecting Weakly Interacting Massive Particles (WIMPs)~\cite{DarkSide50}. The hypothetical WIMPs are expected to be detected through their elastic recoils with ordinary matter via nuclear interactions. Neutrons with energies on the MeV scale, originated by natural uranium and thorium contamination in the detector materials, can mimic the physics signal producing nuclear scatterings in the same energy range of interest of the WIMP search. This neutron-induced irreducible background is the limiting factor for the sensitivity of the current generation of dark matter direct search experiments~\cite{DarkSide20}.

There are various codes specifically developed to calculate neutron yields and energy spectra from ($\alpha$,n) reactions induced by $\alpha$-decay. Some of them are NeuCBOT~\cite{NeuCBOT}, SOURCES~\cite{SOURCES-4C}, NEDIS~\cite{NEDIS} and an application developed by the University of South Dakota (USA), here refered as the USD code~\cite{USD}.

All these codes work in a similar way. They start with a given list of $\alpha$ particle energies or $\alpha$-emitting nuclei. In the second case they use decay data tables to obtain the total amount and energy spectra of  the initial $\alpha$ particles. Then, they use neutron production cross sections and stopping power data tables to calculate the neutron yields, $Y$, given by
\begin{equation}
 Y(E_{\alpha})=\int_{0}^{E_{\alpha}} \frac{\sigma_{(\alpha,Xn)}(E)}{\varepsilon(E)} dE \label{eq:1}
\end{equation}
where $E_{\alpha}$ is the initial energy of the $\alpha$ particle, $\sigma_{(\alpha,Xn)}(E)$ the neutron production cross section, defined as the sum of all the partial cross sections weighted by the number of neutrons produced in each reaction channel, and $\varepsilon(E)$ the stopping cross section of the material considered. The last two variables depend on the $\alpha$-energy $E$. Finally, the neutron energy spectra is obtained either assuming an isotropic neutron angular distribution in the center-of-mass system (SOURCES) or from data tables (rest of the codes).

Until recently, the general-purpose radiation transport simulation codes Geant4~\cite{GEANT4,GEANT4_2016}, standard MCNP~\cite{MCNP61} and FLUKA~\cite{FLUKA} were not able to perform these calculations with the same accuracy. The limitation come from the difficulty of realistically modeling the low energy ($\alpha$,n) reactions. Since version 10.2, released in 2015, Geant4 has incorporated the so called ParticleHP module, which is able to use data libraries originally written in ENDF-6 format~\cite{ENDF6} to manage the non-elastic nuclear reactions of low energy ($<$200 MeV) charged particles~\cite{ParticleHP}. These data libraries contain information describing nuclear interactions, in particular reaction cross sections and secondary particle production. This allows, in principle, to model ($\alpha$,n) reactions with a higher accuracy than that given by the theoretical models implemented in the code. We have investigated the performance of Geant4, using the ParticleHP module, when calculating neutron yields and energy spectra induced by $\alpha$-decay. In this paper we present the results of our study.

The procedure to calculate the neutron production with Geant4 is rather different than in the aforementioned codes. Instead of evaluating the expression given in Eq.~\ref{eq:1} Geant4 performs an explicit transport of the incident $\alpha$ particles through the material. Neutrons are then generated one by one as the nuclear reactions take place. For this reason, a much longer computation time is required by Geant4 than by the other codes, even when biasing techniques are applied. However, this fact is not considered an important issue in most practical cases. On the other hand, Geant4 present some advantages: it allows to integrate in the same simulation both the neutron generation and the neutron transport; it permits to simulate much more complex geometries than the other codes, improving significantly the behaviour at the interfaces between different materials; and, in some specific cases (depending on the information present in the data library used), Geant4 is also able to generate neutrons and $\gamma$-rays in coincidence, as well as other secondary particles emitted in the same nuclear reaction, a fundamental feature in order to fully exploit the tagging of the neutron-induced nuclear recoil.

The calculations presented in this paper are performed with a modified version of geant4.10.4.p01 (version of February 2018), which is able to take into account new features in the incident $\alpha$ data libraries. These modifications will be included in the geant4.10.6 version.

The results of our study are presented organized in three different sections. In Section~\ref{Sec:verification} we show a verification study performed to investigate whether Geant4 provides the results expected from the incident $\alpha$ particle data libraries. Section~\ref{Sec:dataLibs} is devoted to present the available incident-$\alpha$ ENDF-6 format data libraries and the differences among them. Finally, we compare the results obtained with Geant4 and other codes to available experimental data in Section~\ref{Sec:Comparison}.

\section{Verification of Geant4 \label{Sec:verification}}

Three ingredients are needed for the calculation of neutron yields and energy spectra induced by ($\alpha$,n) reactions: (i) a model for simulating the energy loss due to electromagnetic (EM) interactions of the $\alpha$ particles, (ii) nuclear reaction cross sections and (iii) secondary-neutron energy and angular distributions. We have built a Geant4 application which uses the EM processes implemented in Geant4 for simulating the energy loss, and neutron production cross sections and secondary particle production from ENDF-6 format data libraries. The aim of our study is to verify that the neutron yields and energy spectra produced by Geant4 are in agreement with the information present in the libraries.

In this application, $\alpha$ particles of a given initial energy are transported inside a single volume, made of a given material, large enough so they do not escape. Two histograms are generated in each simulation: one with the secondary neutron energy spectrum and the other with the $\alpha$ particle flux inside the volume. Secondary particles are killed when created, to ensure that all neutrons are created by ($\alpha$,Xn) reactions. We use the G4EmStandardPhysics\_option4 package for modeling the EM processes, which takes the stopping powers from the ICRU 49 report~\cite{ICRU49} below 8 MeV and the Bethe-Bloch formula above. To control the level of detail in which the transport of the $\alpha$ particles is carried out, we use the G4UserLimits class to define a maximum allowed step length, $S_{max}$. Elastic and non-elastic nuclear interactions are modeled with the G4HadronElasticPhysics and the G4ParticleHP packages, respectively. Various ENDF-6 format incident $\alpha$ particle data libraries are converted into the Geant4 format with the same tool~\cite{IEEEG4NDL} used to generate the neutron data libraries available in~\cite{NeutronLibsWeb}. 

Since EM processes dominate over nuclear ones, it is necessary to use biasing techniques to reduce computation times. We use the generic biasing scheme provided by Geant4 with the same implementation that is found in the \textit{extended/biasing/GB01} example distributed with the source code. In our application, nuclear cross sections are enhanced by a given factor, hereafter referred to as \textit{bias factor}, $F_{B}$. Secondary neutrons are then generated with statistical weights which take into account this enhancement. If the bias factor is large enough, the weight of the initial $\alpha$ particle is increased along the tracking process to correct for the self shielding effects caused by the cross section biasing.

We have preformed simulations with different materials, nuclear data libraries, bias factors and maximum allowed step lengths. To verify our results, we compare the yields obtained by tallying the secondary neutrons ($Y$) with the neutrons resulting from the convolution of the $\alpha$ particle flux with the neutron production cross section ($Y_{r}$),

\begin{equation}
 Y_{r}=V\cdot N\cdot\int \Phi_{\alpha}(E)\sigma_{(\alpha,Xn)}(E) dE \label{eq:2}
\end{equation}
where $V$ is the volume of the material, $N$ the atom density and $\Phi_{\alpha}(E)$ the $\alpha$ particle differential flux. We present some results to illustrate the performance of the application program in Tables~\ref{tab:Comp-Bias01} and~\ref{tab:Comp-StepMax}.

\begin{table}[ht]
\centering\
\begin{tabular}{cccccc}
\hline
$F_{B}$ & $n_{n}$ & $Y$ & $\Delta Y$(\%) & $({n_{n}})^{-1/2}$(\%) & $Y/Y_{r}$ \\  
\hline
1        &  1021    & 102.1 & 3.1    & 3.1    &   0.931  \\
10       &  10787   & 107.9 & 0.96   & 0.96   &   0.980  \\
10$^{2}$ &  108369  & 109.0 & 0.30   & 0.30   &   0.994  \\
10$^{3}$ &  1034886 & 109.2 & 0.098  & 0.098  &   0.997  \\
10$^{4}$ &  6642063 & 109.2 & 0.041  & 0.039  &   0.996  \\
10$^{5}$ &  9998626 & 108.2 & 0.67   & 0.033  &   0.987  \\
10$^{6}$ &  9997953 & 25.3  & 21     & 0.032  &   0.230  \\
\hline
\end{tabular}
\caption{Results obtained from simulations of 10$^{7}$ $\alpha$ particles of 10 MeV inside $^{13}$C, for different bias factors, $F_{B}$ (see text for details).}
\label{tab:Comp-Bias01}
\end{table}

Table~\ref{tab:Comp-Bias01} shows the results of several simulations performed with different bias factors. In all the cases, 10 MeV initial $\alpha$ particles are transported inside a $^{13}$C volume. A detailed tracking of the $\alpha$ particles is performed, using a maximum allowed step length of 0.1 $\mu$m. To model the nuclear interactions the JENDL/AN-2005~\cite{JENDL/AN-2005} data library is used. Each row corresponds to a different simulation, performed with the bias factor given in the first column. $n_{n}$ is the total number of neutrons generated by 10$^{7}$ initial $\alpha$ particles. $Y$ and $\Delta Y$ are, respectively, the neutron yield and its uncertainty due to counting statistics, obtained by tallying the secondary neutrons. The fifth column represents the uncertainty in the yield  due to counting statistics, assuming the weights of all the neutrons are equal. Finally, the last column compares $Y$ and $Y_{r}$. Since the uncertainty in $Y_{r}$ due to counting statistics is very small compared to $Y$, the uncertainty values given in the fourth column are also valid for $Y/Y_{r}$.

As expected, the number of secondary neutrons generated in the simulation increases with $F_{B}$. This number saturates at some point between $F_{B}=10^{4}$ and $F_{B}=10^{5}$, where in almost every event a neutron is generated. All the values of $Y$ are consistent with $Y_{r}$ but the one obtained with $F_{B}=10^{6}$, a bias factor at least one order of magnitude above the saturation value. The results show that, for reasonably large values of $F_{B}$, below the saturation point, the amount of neutrons generated by Geant4 is in agreement with the expectations for the the cross sections in the nuclear data libraries. Similar results are obtained for other target nuclei, thus verifying in part the performance of the G4ParticleHP package and the biasing process. 

The uncertainty in the obtained yields decreases with $F_{B}$ until reaching values close to saturation ($F_{B}=10^{4}$), and then it begins to increase. From the comparison between $\Delta Y$ and $({n_{n}})^{-1/2}$, it follows that up to $F_{B}=10^{3}$ all the neutrons generated have practically the same weight, i.e. the weight of the $\alpha$ particle do not change along the track. Above $F_{B}=10^{4}$, $\Delta Y$ and $({n_{n}})^{-1/2}$ have no longer the same value, indicating that the weight of the $\alpha$ particle increases along the track to account for self shielding effects due to the cross section enhancement.

\begin{table}[ht]
\centering\
\begin{tabular}{ccccc}
\hline
              & \multicolumn{2}{c}{$^{13}$C} & \multicolumn{2}{c}{$^{14}$N} \\\cmidrule{2-3} \cmidrule{4-5}
$S_{max}$(mm) & Time (s) & $Y/Y_{r}$ & Time (s) & $Y/Y_{r}$ \\  
\hline
10$^{-1}$ & 8.8$\times$10$^{2}$ & 0.955(1) & 7.3$\times$10$^{2}$ & 0.876(2) \\
10$^{-2}$ & 1.8$\times$10$^{3}$ & 0.947(1) & 1.8$\times$10$^{3}$ & 0.927(2) \\
10$^{-3}$ & 1.6$\times$10$^{4}$ & 0.985(1) & 1.5$\times$10$^{4}$ & 0.987(2) \\
10$^{-4}$ & 1.5$\times$10$^{5}$ & 0.997(1) & 1.4$\times$10$^{5}$ & 0.995(2) \\
\hline
\end{tabular}
\caption{Computation time and $Y/Y_{r}$ ratios obtained from simulations of 10$^{7}$ $\alpha$ particles of 10 MeV inside $^{13}$C and $^{14}$N, using JENDL/AN-2005 and a biasing factor of 10$^{3}$, for different maximum allowed step lengths $S_{max}$.}
\label{tab:Comp-StepMax}
\end{table}

Table~\ref{tab:Comp-StepMax} displays the effect of varying the maximum allowed step length. The smaller the value of $S_{max}$, the more detailed the tracking of the $\alpha$ particle is, which produces more precise yields, but requires longer computation times. Indeed, there is an almost linear relationship between $S_{max}$ and computation times. The best results are obtained with $S_{max}=10^{-4}$ mm. However, the deviations in the yields obtained with larger $S_{max}$ values may not be important in practice for many applications. Results do not improve with values of $S_{max}$ lower than 10$^{-4}$ mm.

\begin{figure}[ht]
\begin{center}
\includegraphics[width=0.45\linewidth]{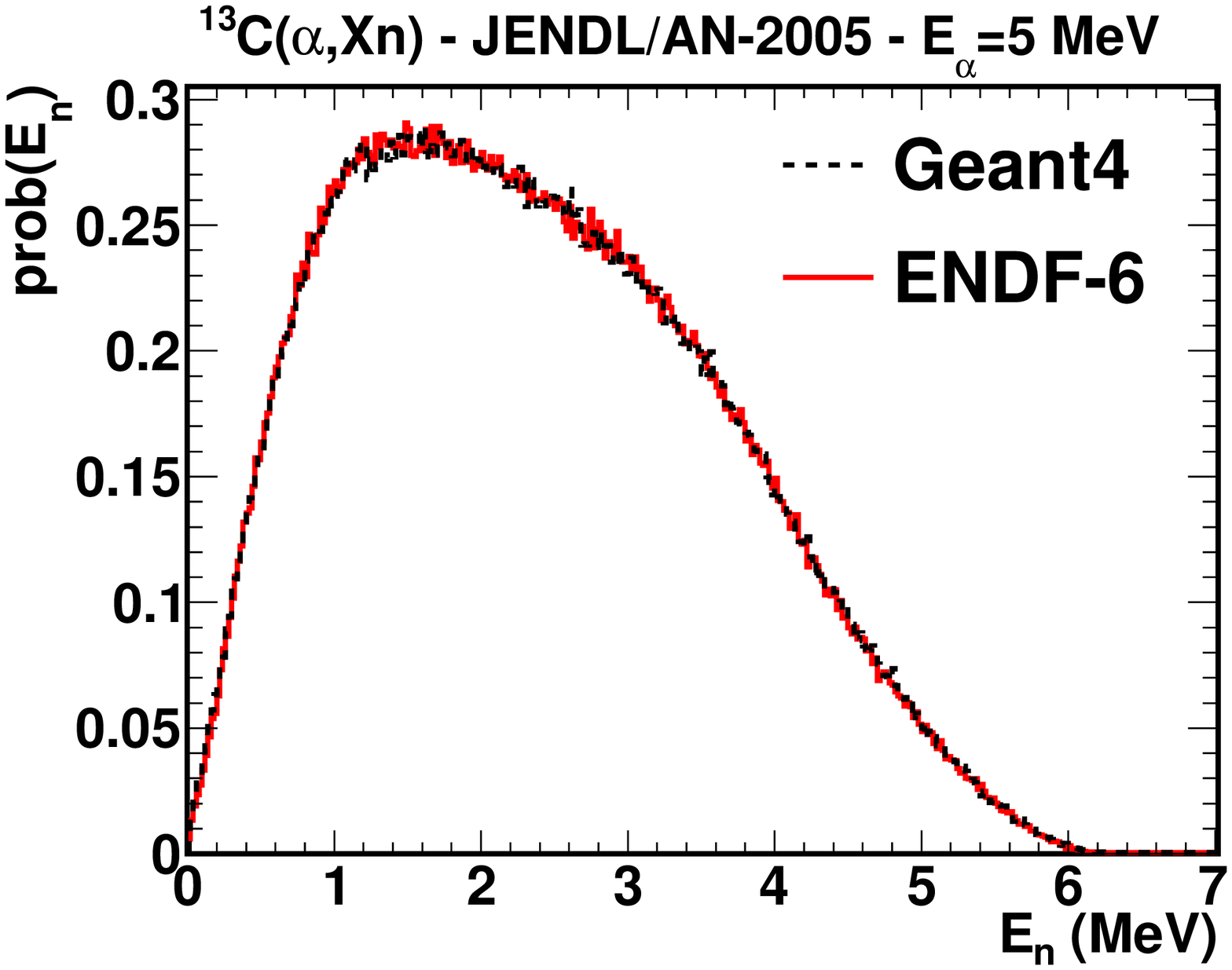}
\includegraphics[width=0.45\linewidth]{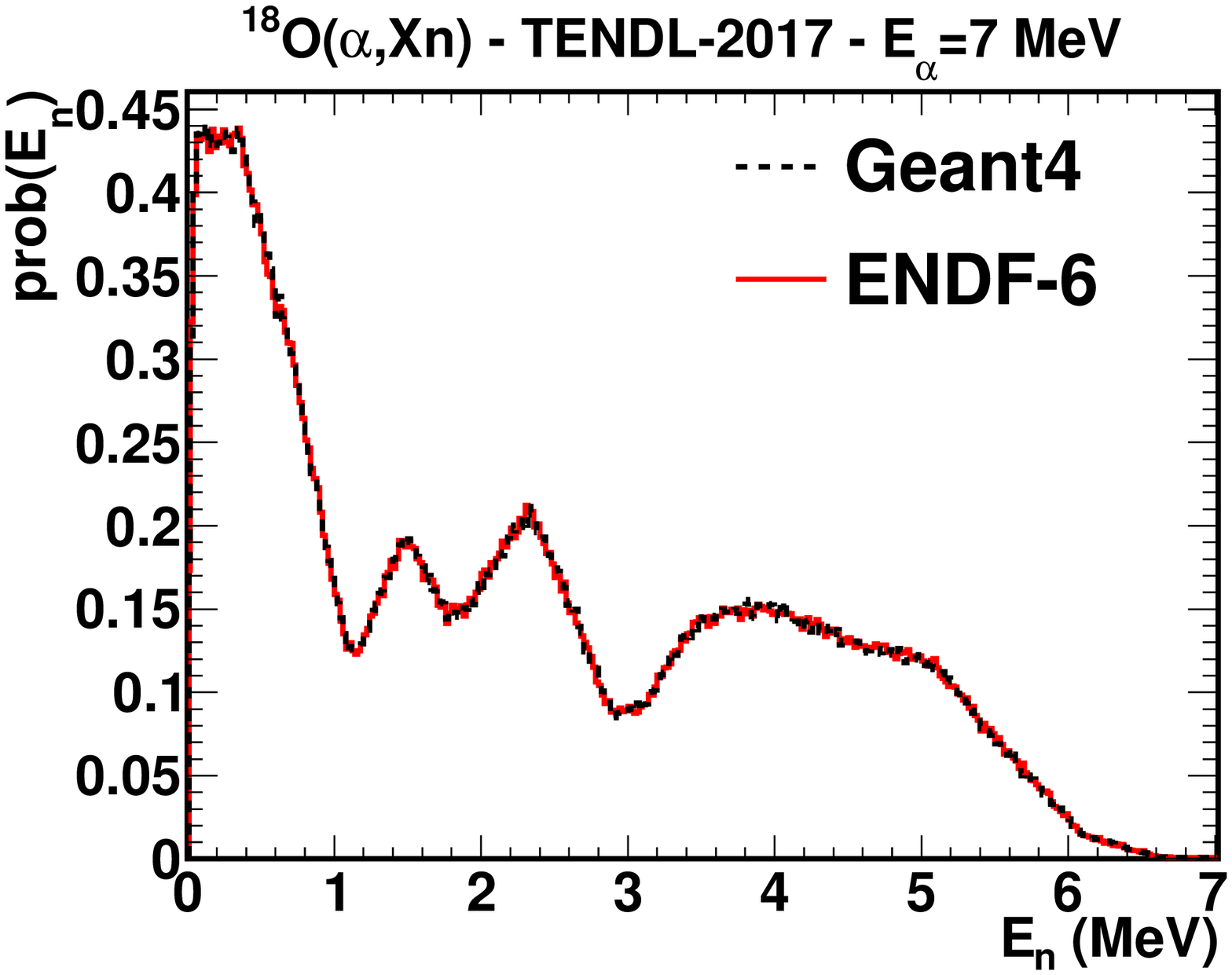}
\caption{Neutron energy spectra from ($\alpha$,Xn) reactions in $^{13}$C (top) and $^{18}$O (bottom) at two fixed $\alpha$ particle energies, 5 and 7 MeV, respectively. The spectra obtained with Geant4 are compared with the spectra present in the ENDF-6 nuclear data libraries used in each case.}
\label{fig:NeuSpecVerification}
\end{center}
\end{figure}

To verify the correctness of the neutron energy spectra, we compare several spectra sampled by Geant4 at a fixed $\alpha$ particle energy with the original data present in the ENDF-6 format files. The Geant4 spectra are calculated with the \textit{extended/hadronic/Hadr03} example distributed with the source code. In this example, the spectra are obtained from Monte Carlo simulations, forcing collisions of an incident $\alpha$ particle at a fixed energy on a given material, and recording the energies of the resulting secondary neutrons. The ENDF-6 spectra are extracted from a program developed by the authors~\cite{IEEEG4NDL},that reads the ENDF-6 data libraries. The comparison of the spectra for various nuclei at several $\alpha$ particle energies displays an excellent agreement among the results of both methods. Two examples are presented in figure~\ref{fig:NeuSpecVerification}.

\FloatBarrier

\section{Alpha incident data libraries \label{Sec:dataLibs}}

At present, JENDL/AN-2005~\cite{JENDL/AN-2005} is the only publicly available evaluated ENDF-6 format library for $\alpha$ incident particles~\cite{IAEA_ENDF}. It contains neutron production cross sections and secondary neutron energy and angular distributions for the following isotopes: $^{6,7}$Li, $^{9}$Be, $^{10,11}$B,$^{12,13}$C, $^{14,15}$N, $^{17,18}$O, $^{19}$F, $^{23}$Na, $^{27}$Al and $^{28,29,30}$Si. In this context, the term \textit{evaluated} means that experienced nuclear physicists has analyzed in detail the experimental data available, combining that information with theoretical models, performing a detailed study for each isotope.

Other (non-evaluated) ENDF-6 format data libraries are the TENDL~\cite{TENDL}, which have been created from the output of the TALYS code~\cite{TALYS} calculations. Three different versions of the TENDL libraries have been considered in our study: TENDL-2014~\cite{TENDL2014}, TENDL-2015~\cite{TENDL2015} and TENDL-2017~\cite{TENDL2017}. In the three cases, there are two different sub-versions of the library. In one of them all the non-elastic reactions are grouped together into a single channel. A total non-elastic cross section is provided together with the particle yield and energy-angular distribution of every secondary particle. The other sub-version contains explicit cross sections up to 30 MeV.

\begin{figure}[ht]
\begin{center}
\includegraphics[width=0.5\linewidth]{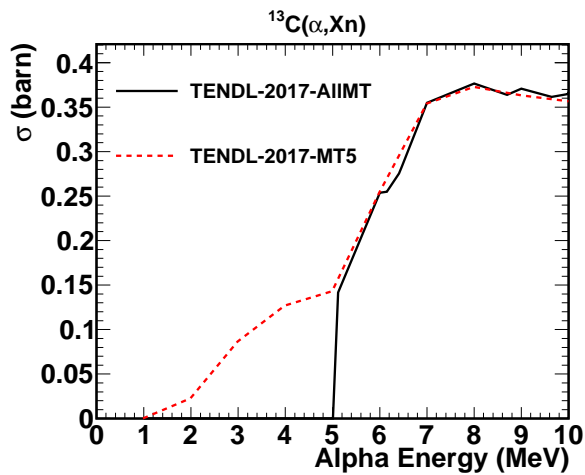}
\caption{Neutron production cross sections of $^{13}$C in the two sub-versions of the TENDL-2017 library, one including all the explicit reaction channels up to 30 MeV (TENDL-2017-AllMT) and the other with all the non-elastic channels grouped together (TENDL-2017-MT5).}
\label{fig:MT5_VS_AllMT}
\end{center}
\end{figure}

Having explicit reaction channels is a big advantage for some applications. It allows, for example, to generate neutrons in coincidence with $\gamma$-rays, which is crucial in calculations related to low background experiments. When all the non-elastic reactions appear together in a single channel, the information concerning secondary particle production is given as independent particle yields, without providing any correlation in the production of the different secondary particles. Unfortunately, we have found that the TENDL sub-versions that include explicit reaction channels at low energies have not been constructed correctly for some isotopes. One example is given in figure~\ref{fig:MT5_VS_AllMT}, where the two sub-versions of the $^{13}$C($\alpha$,Xn) cross section in TENDL-2017 are plotted together. Contrary to the expectations, they differ significantly, due to a very high threshold in the sub-version with explicit channels. Note that the $^{13}$C($\alpha$,n)$^{16}$O reaction has no threshold. A similar situation can be found for TENDL-2014 and TENDL-2015, and for other isotopes such as $^{11}$B, $^{14,15}$N or $^{29,30}$Si. For this reason, we have excluded the TENDL sub-versions with explicit reaction channels from our study. In the rest of this document, whenever we refer to a TENDL library, we mean the sub-version with all the non-elastic channels together.

\begin{figure*}[htb]
\begin{center}$
\begin{array}{c c c}
\includegraphics[width=0.33\linewidth]{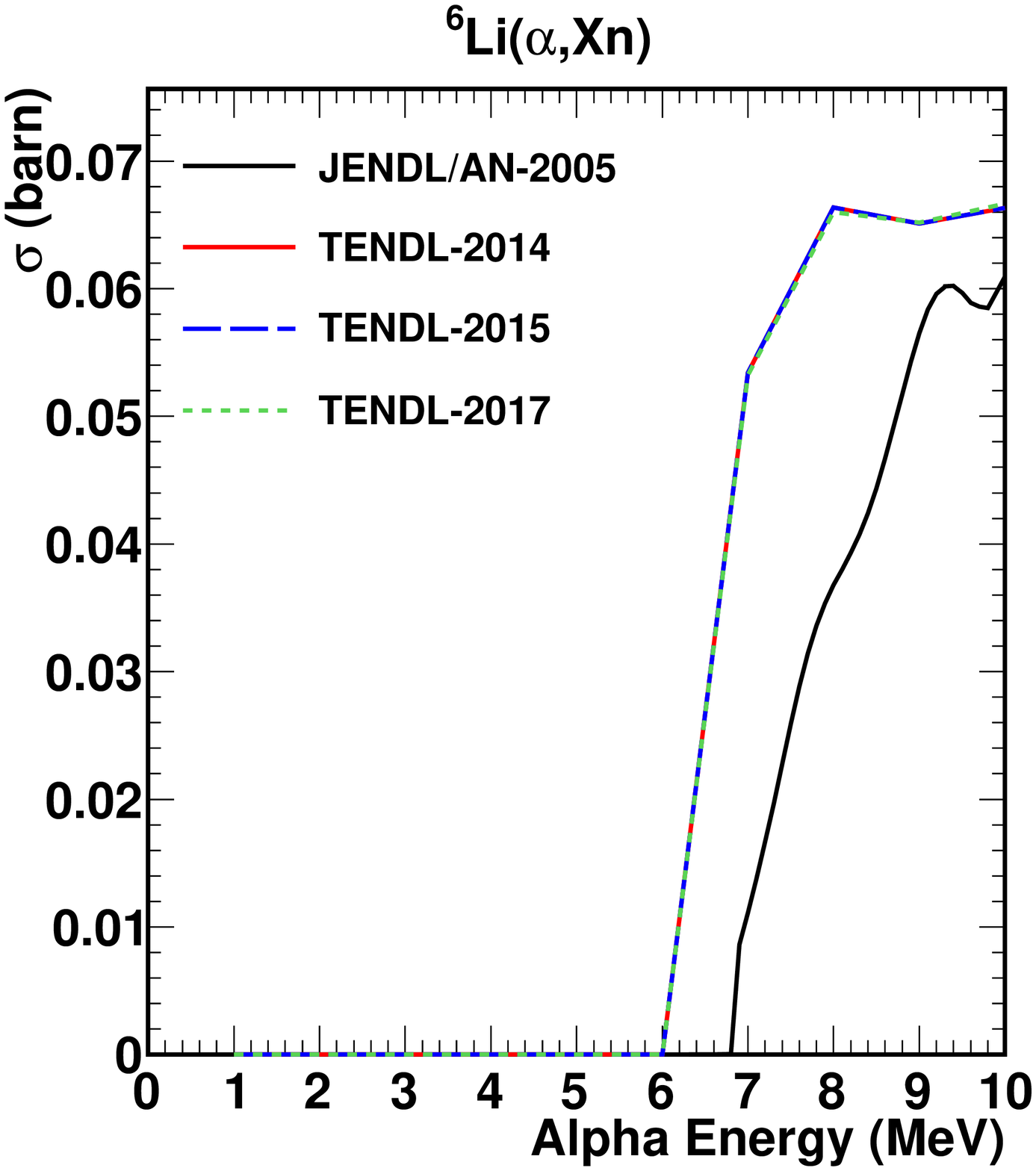} &
\includegraphics[width=0.33\linewidth]{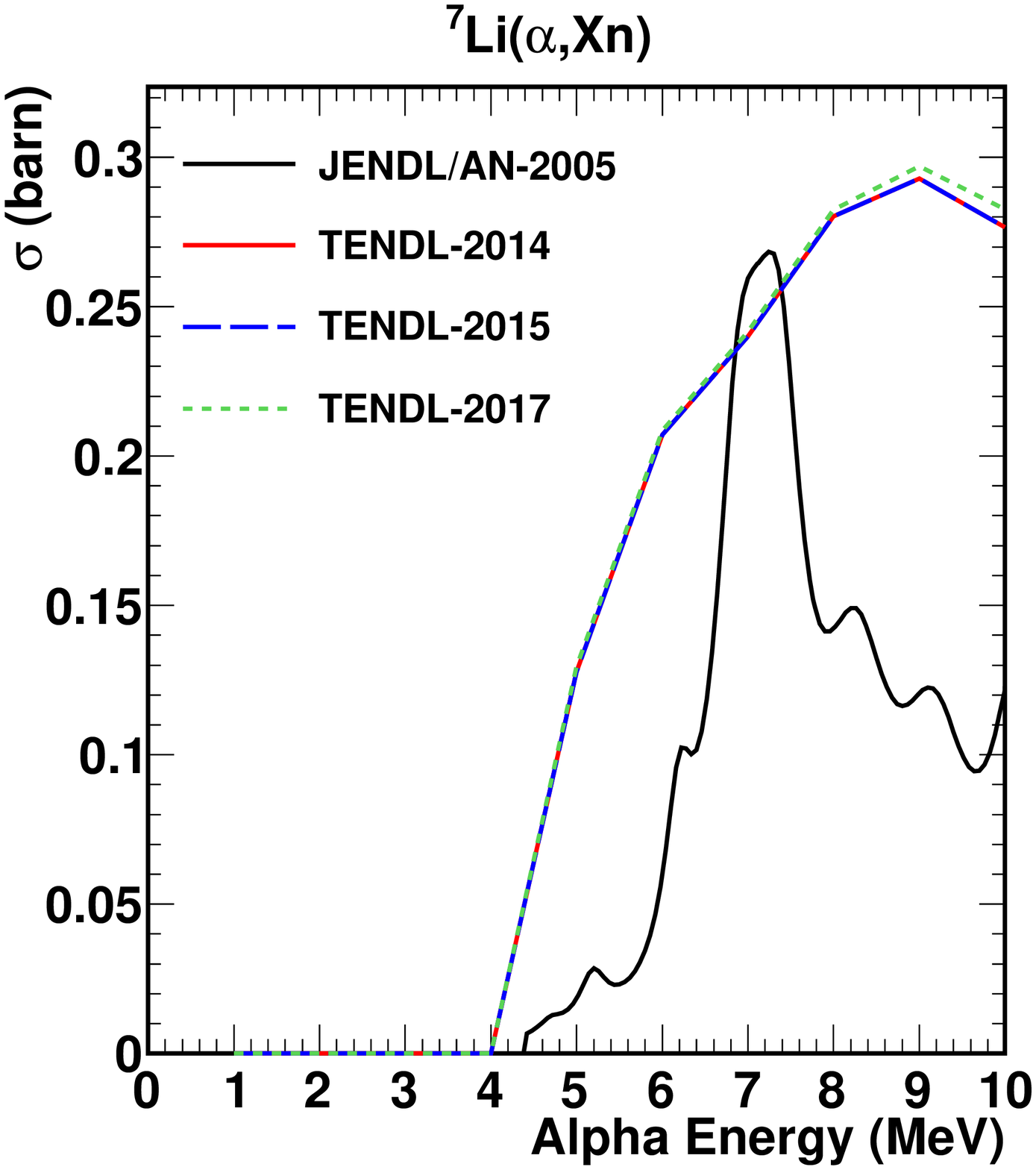} &
\includegraphics[width=0.33\linewidth]{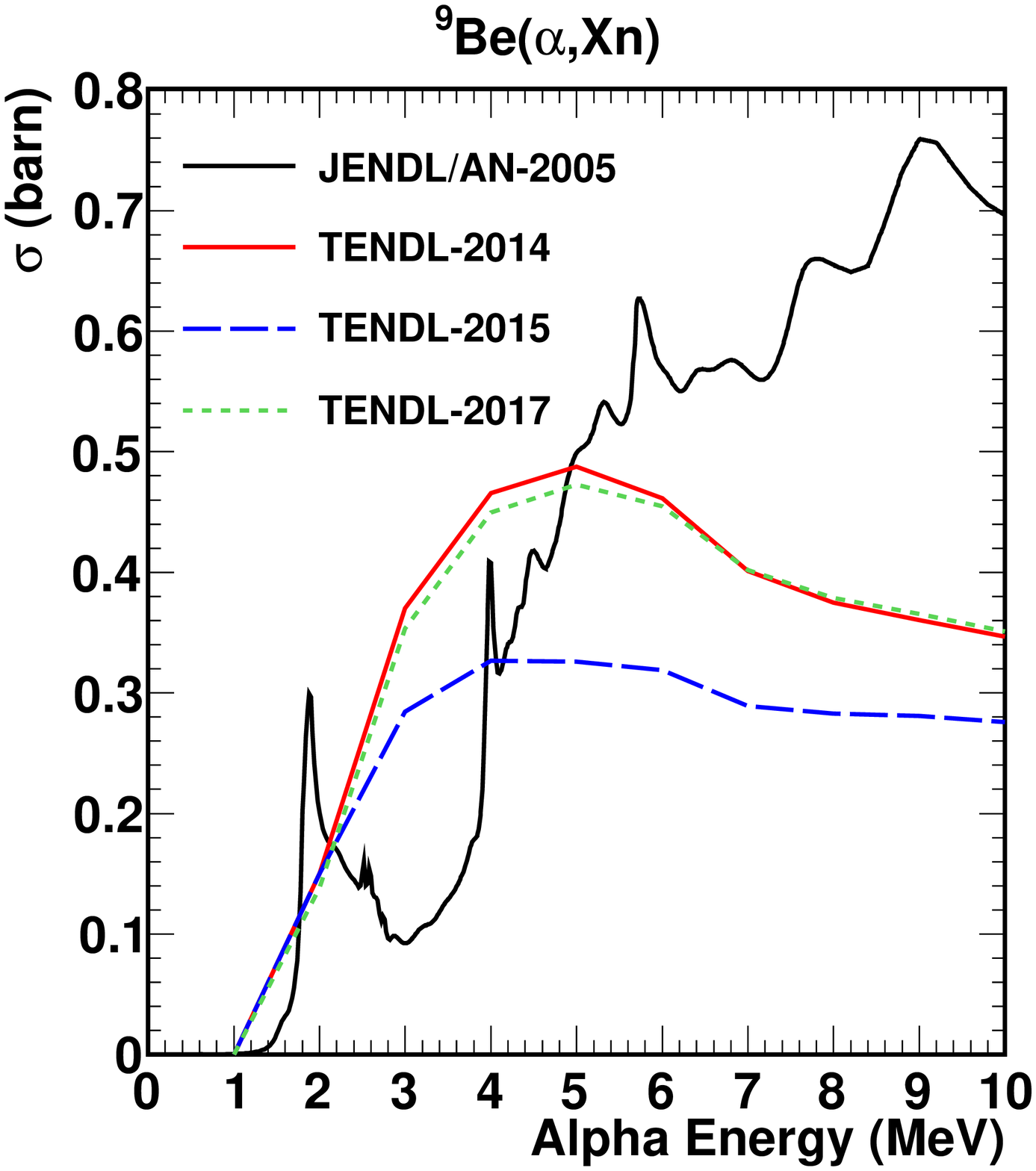} \\
\includegraphics[width=0.33\linewidth]{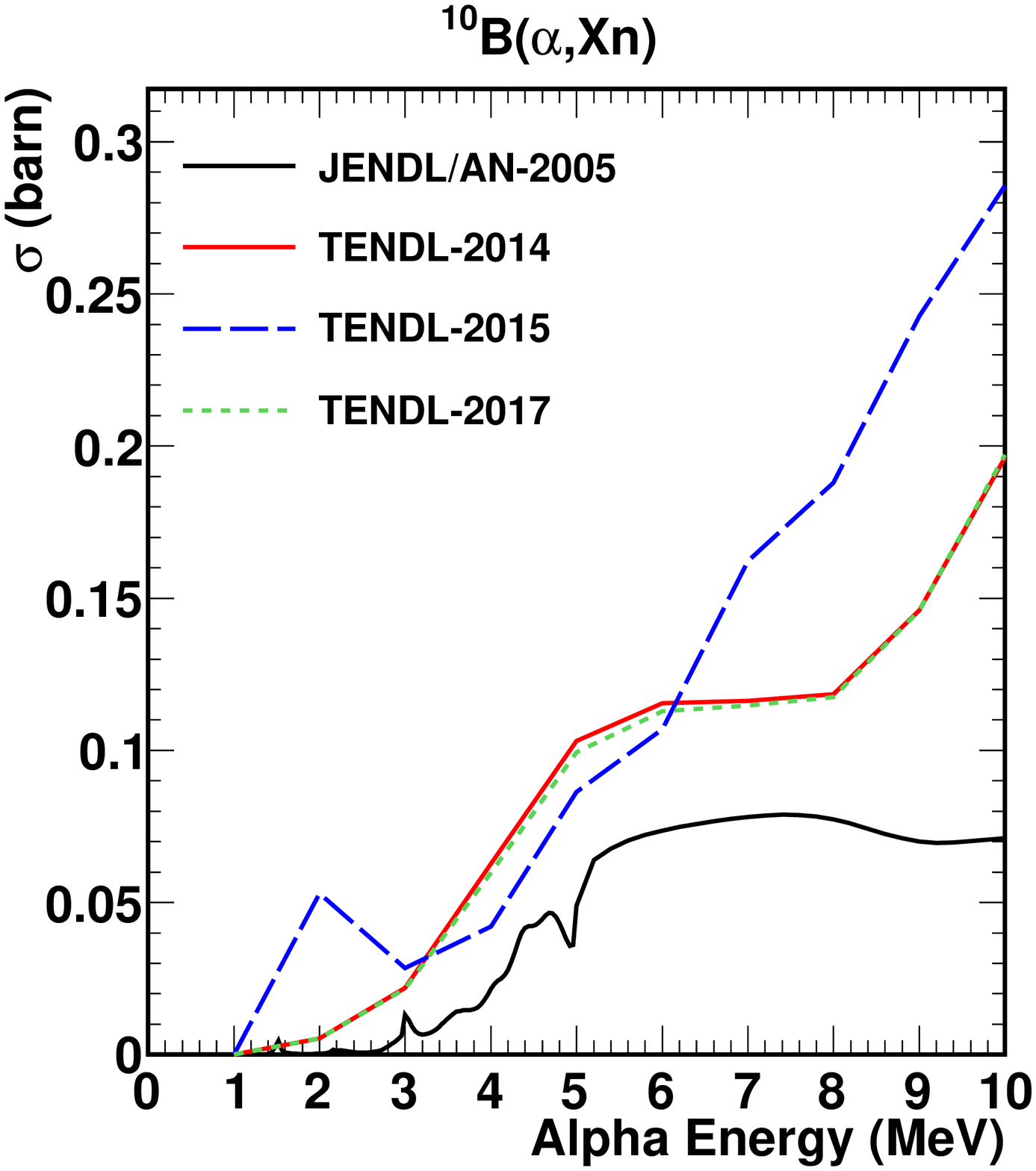} &
\includegraphics[width=0.33\linewidth]{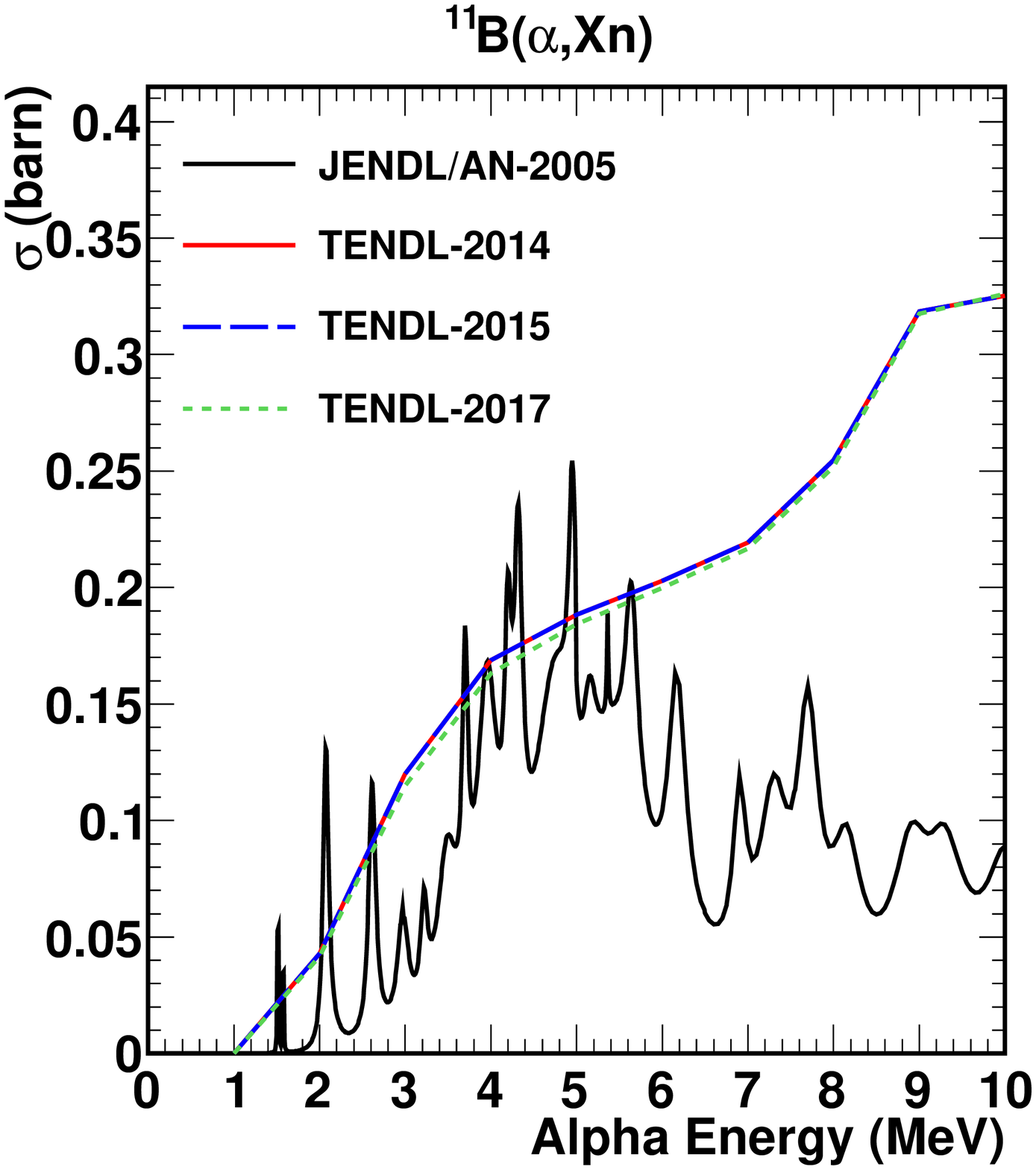} &
\includegraphics[width=0.33\linewidth]{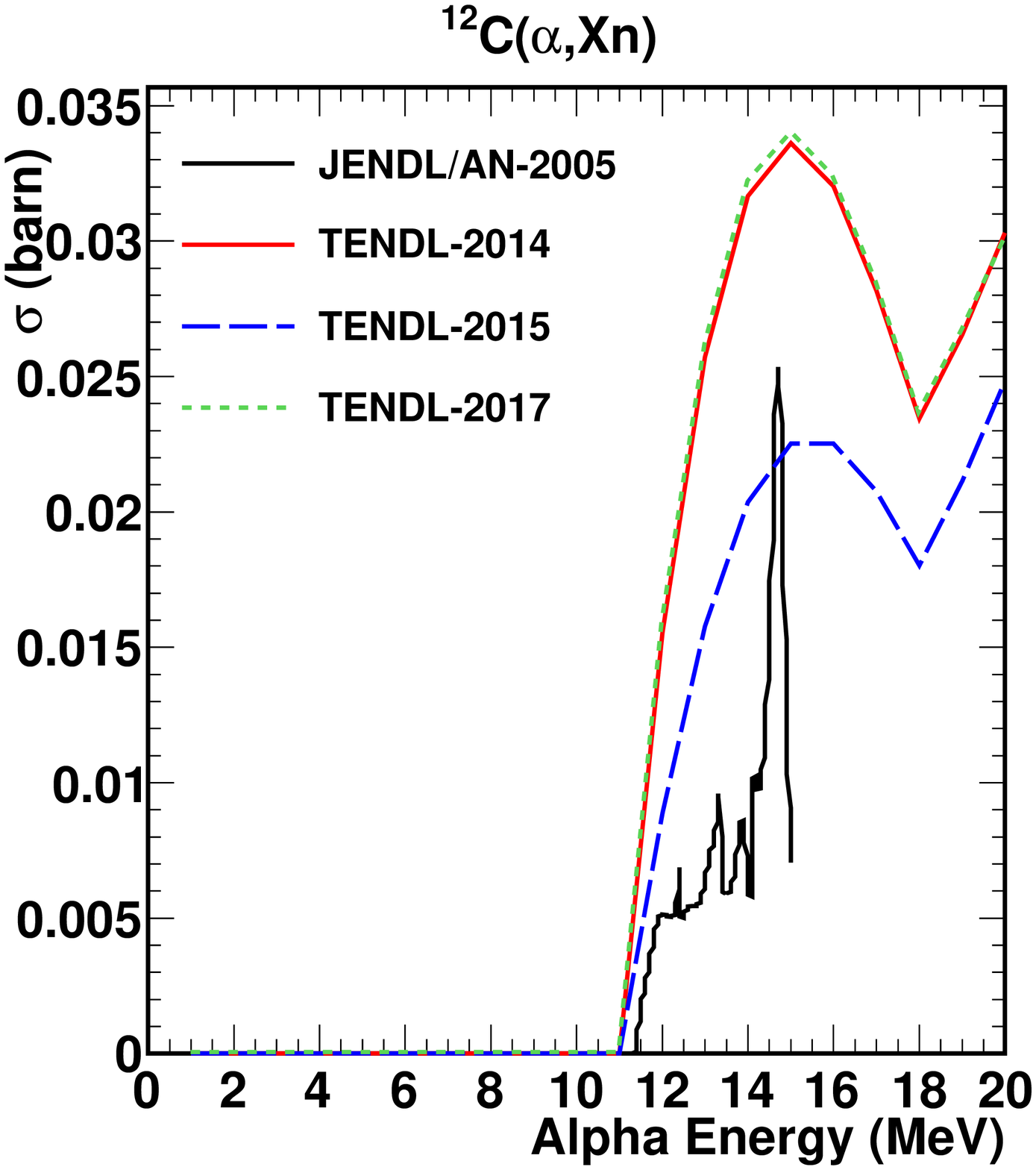} \\
\includegraphics[width=0.33\linewidth]{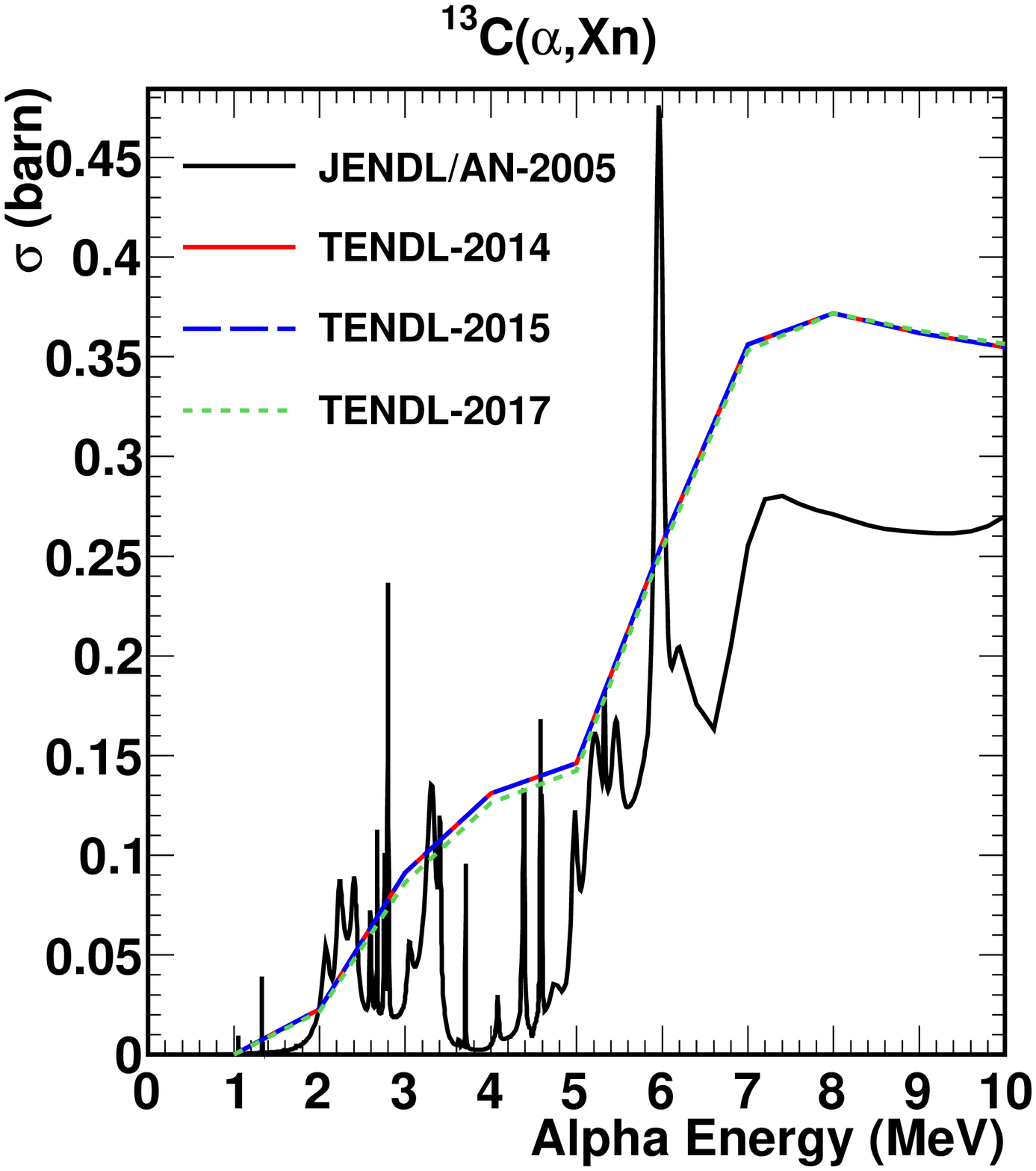} &
\includegraphics[width=0.33\linewidth]{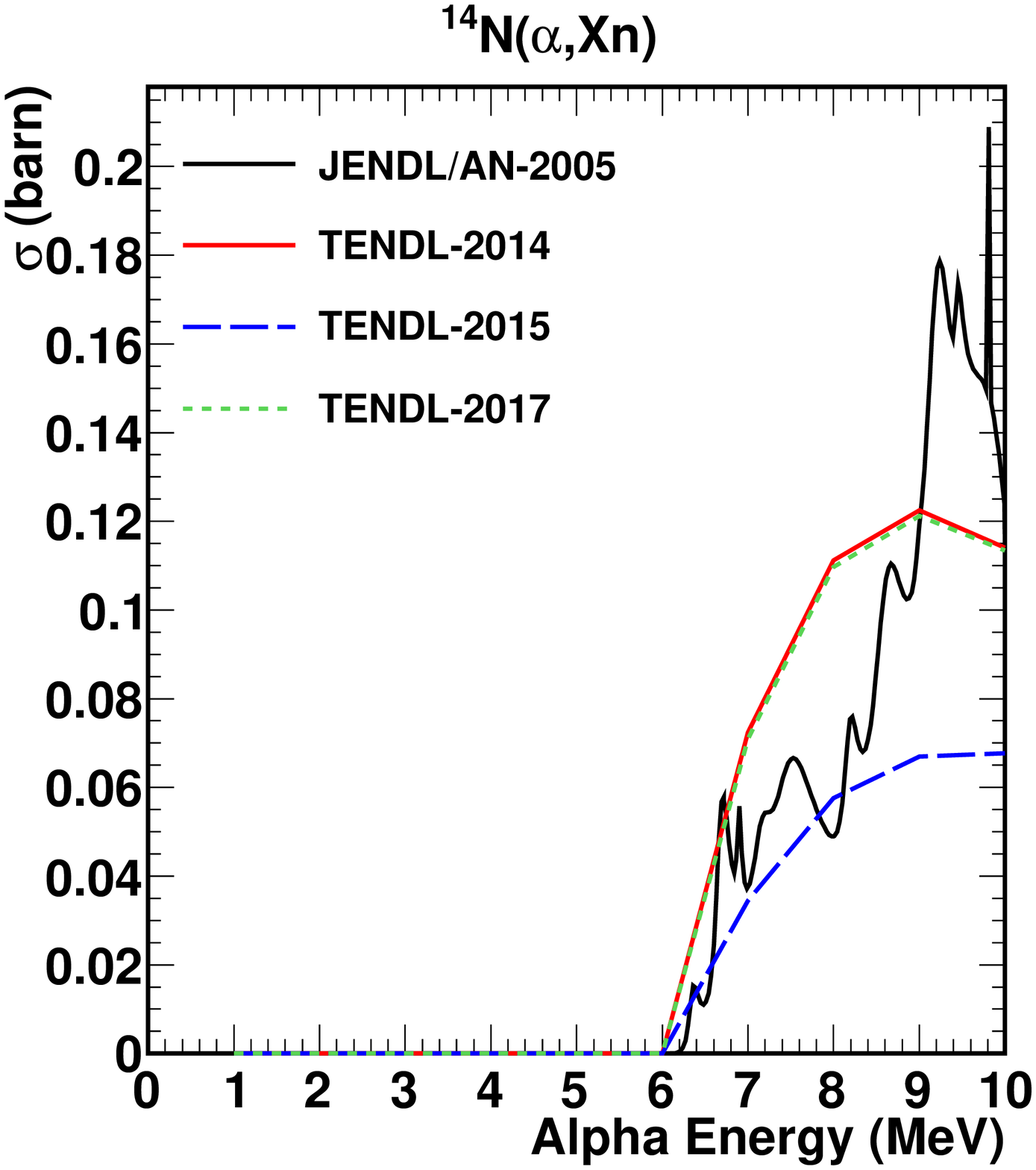} &
\includegraphics[width=0.33\linewidth]{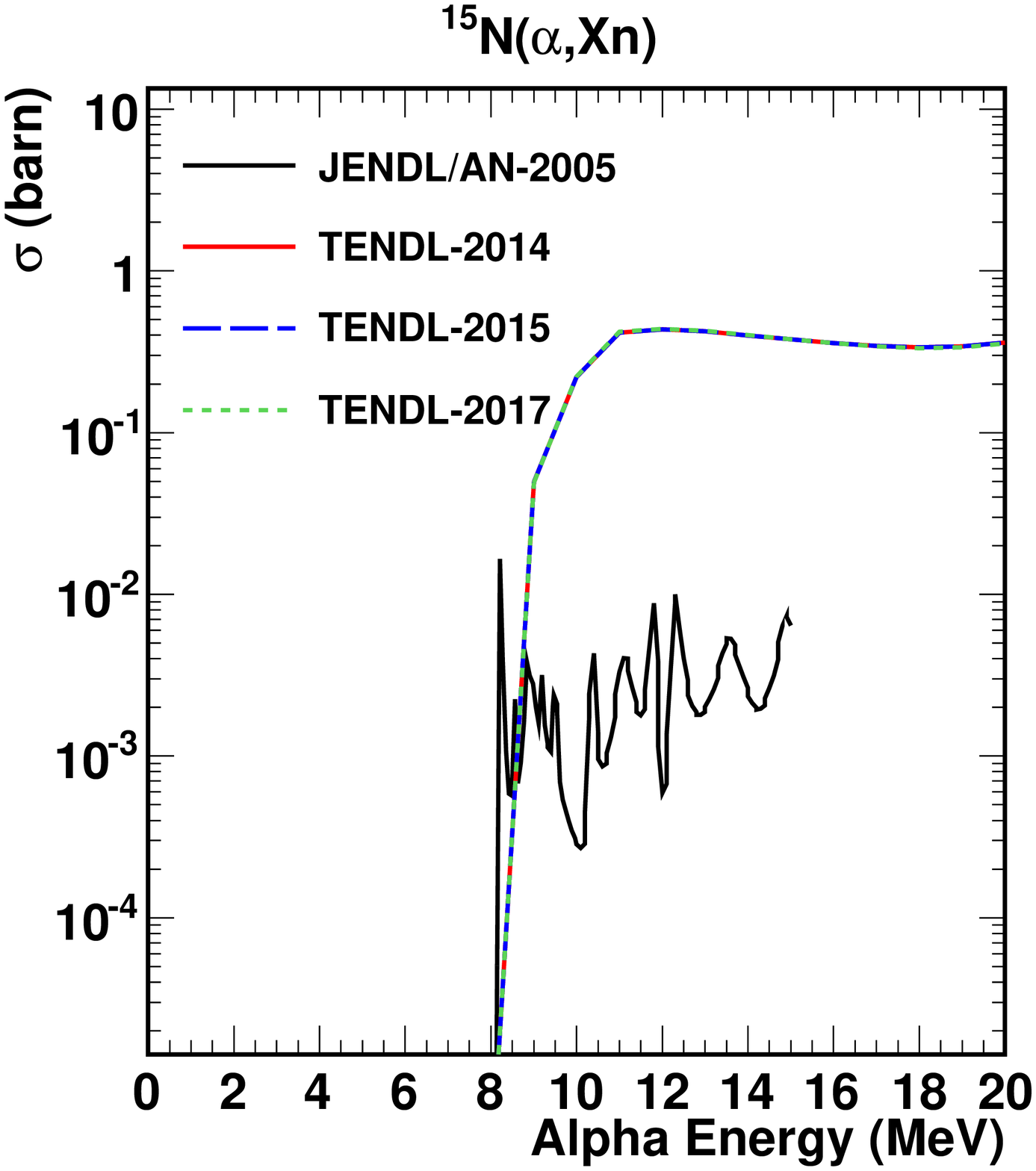} \\
\end{array}$
\end{center}
\caption{Neutron production cross sections at low $\alpha$ particle energies for different isotopes and libraries.}
\label{fig:XS_01}
\end{figure*}

\begin{figure*}[htb]
\begin{center}$
\begin{array}{c c c}
\includegraphics[width=0.33\linewidth]{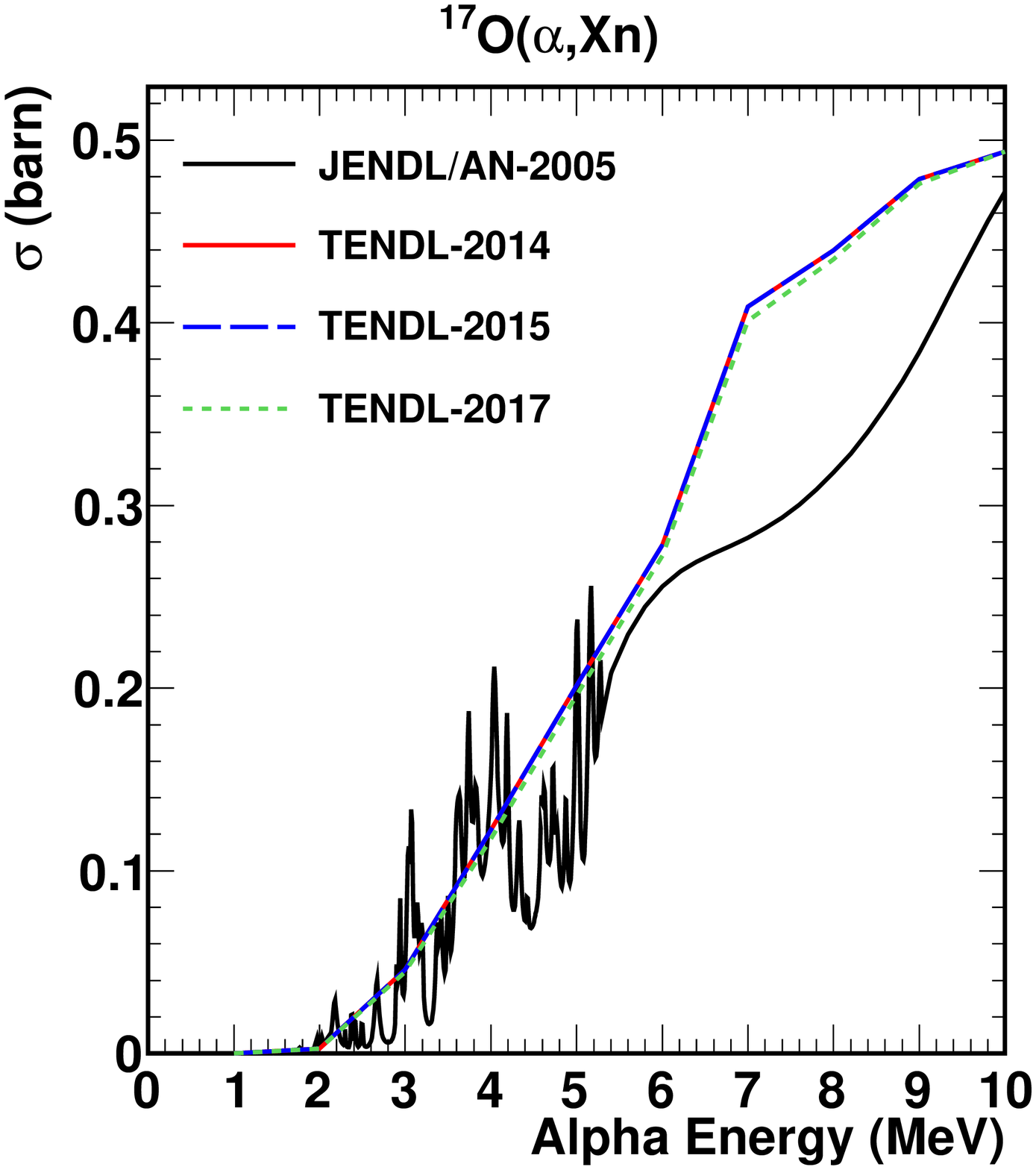} &
\includegraphics[width=0.33\linewidth]{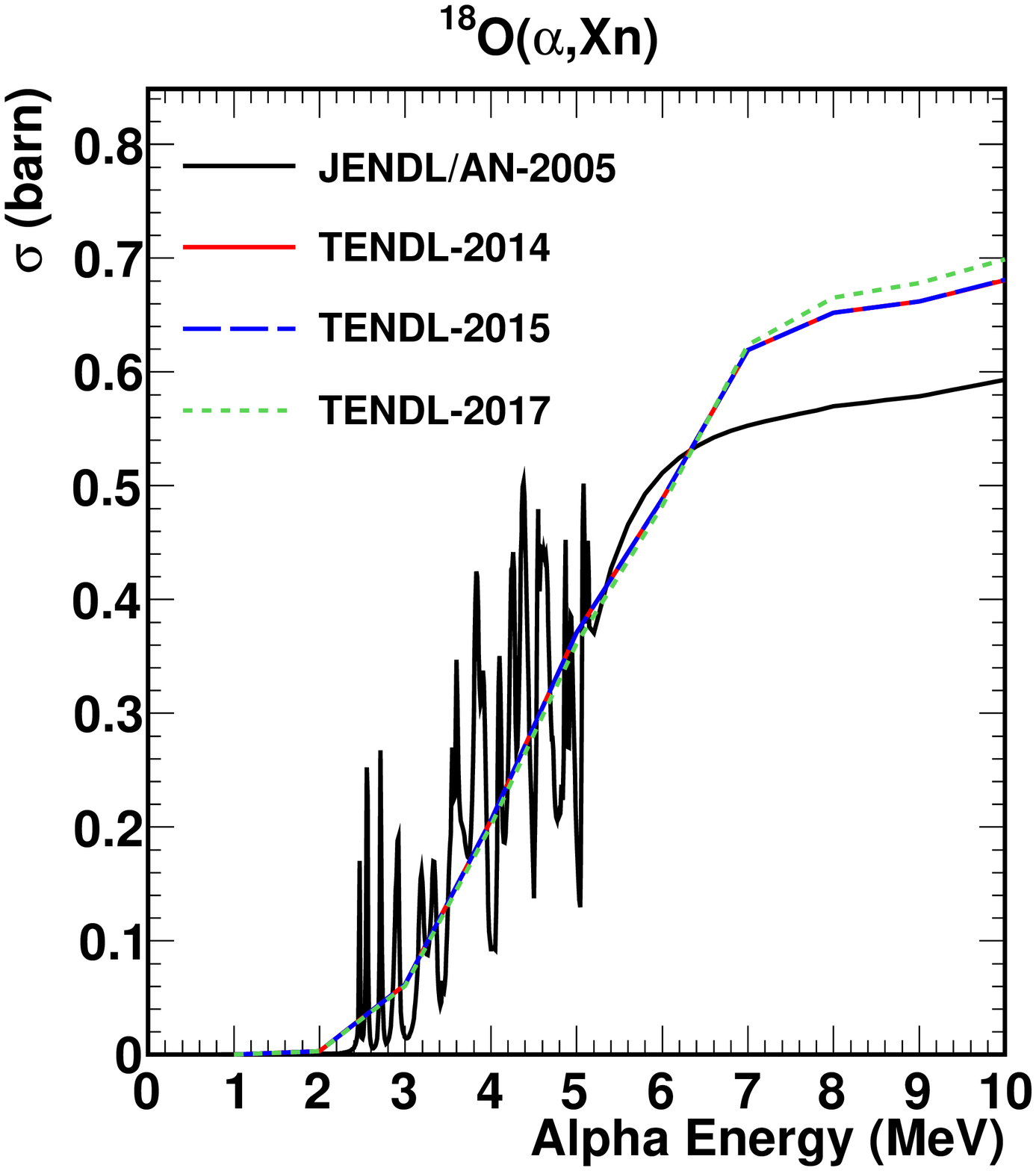} &
\includegraphics[width=0.33\linewidth]{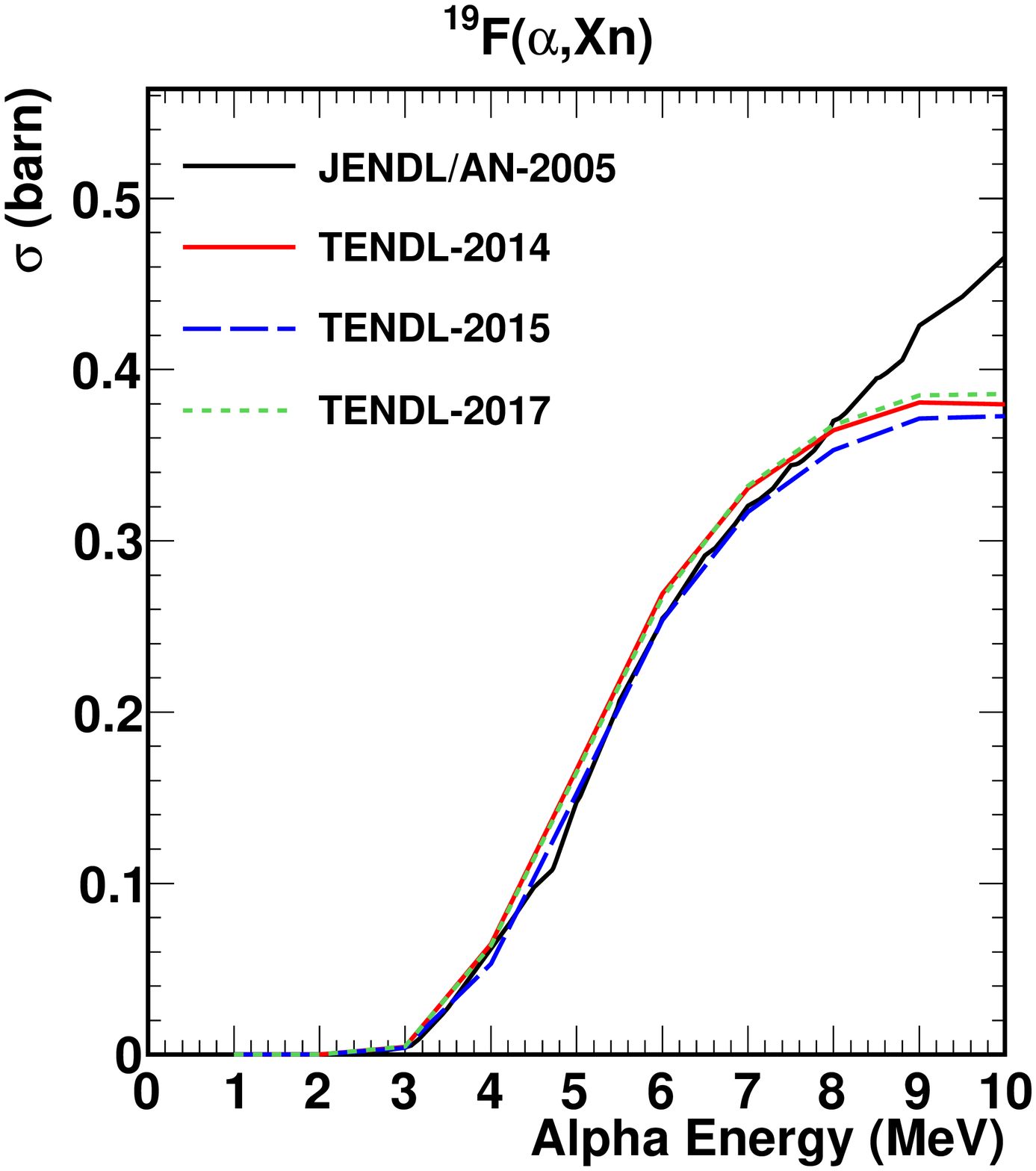} \\
\includegraphics[width=0.33\linewidth]{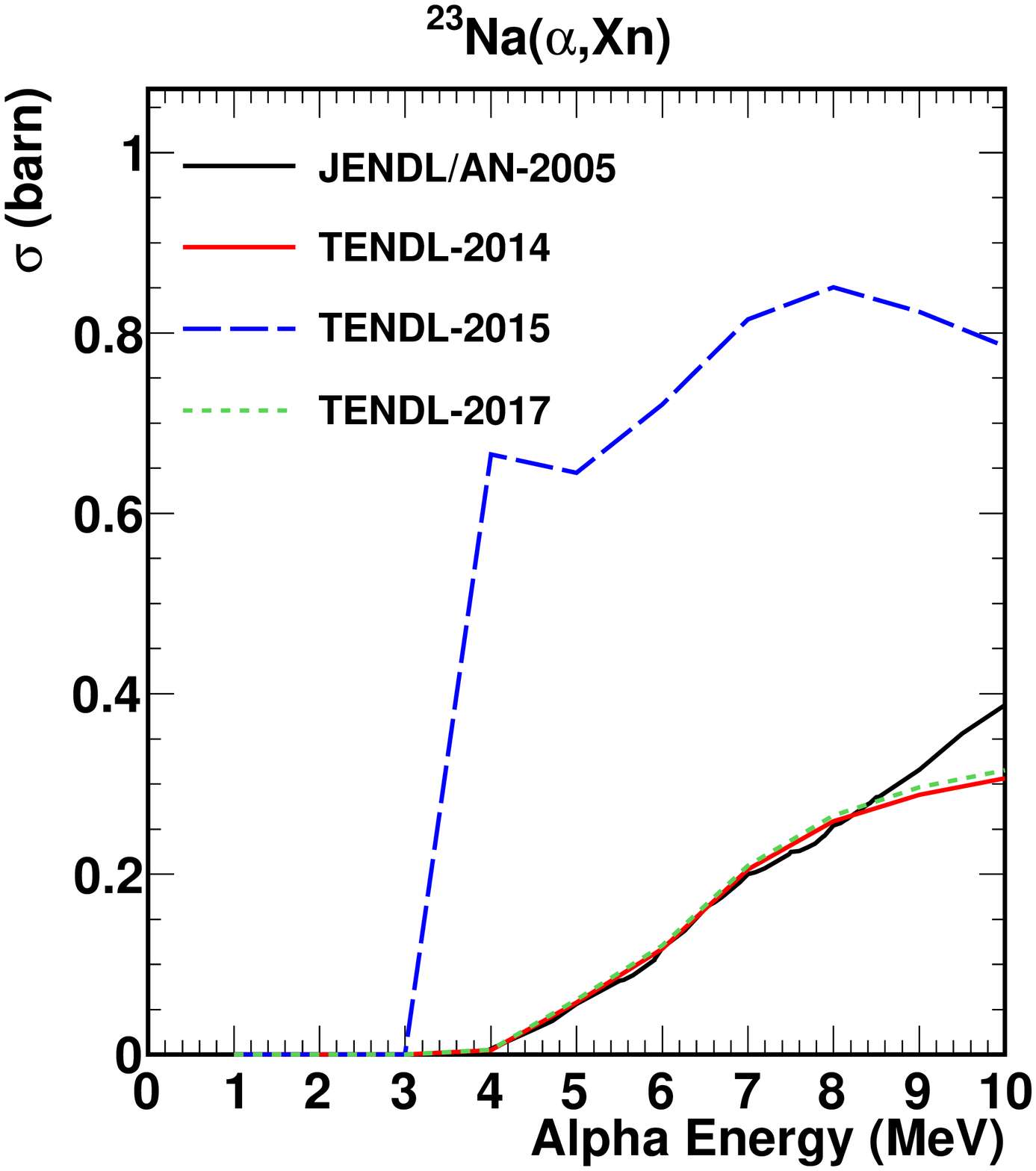} &
\includegraphics[width=0.33\linewidth]{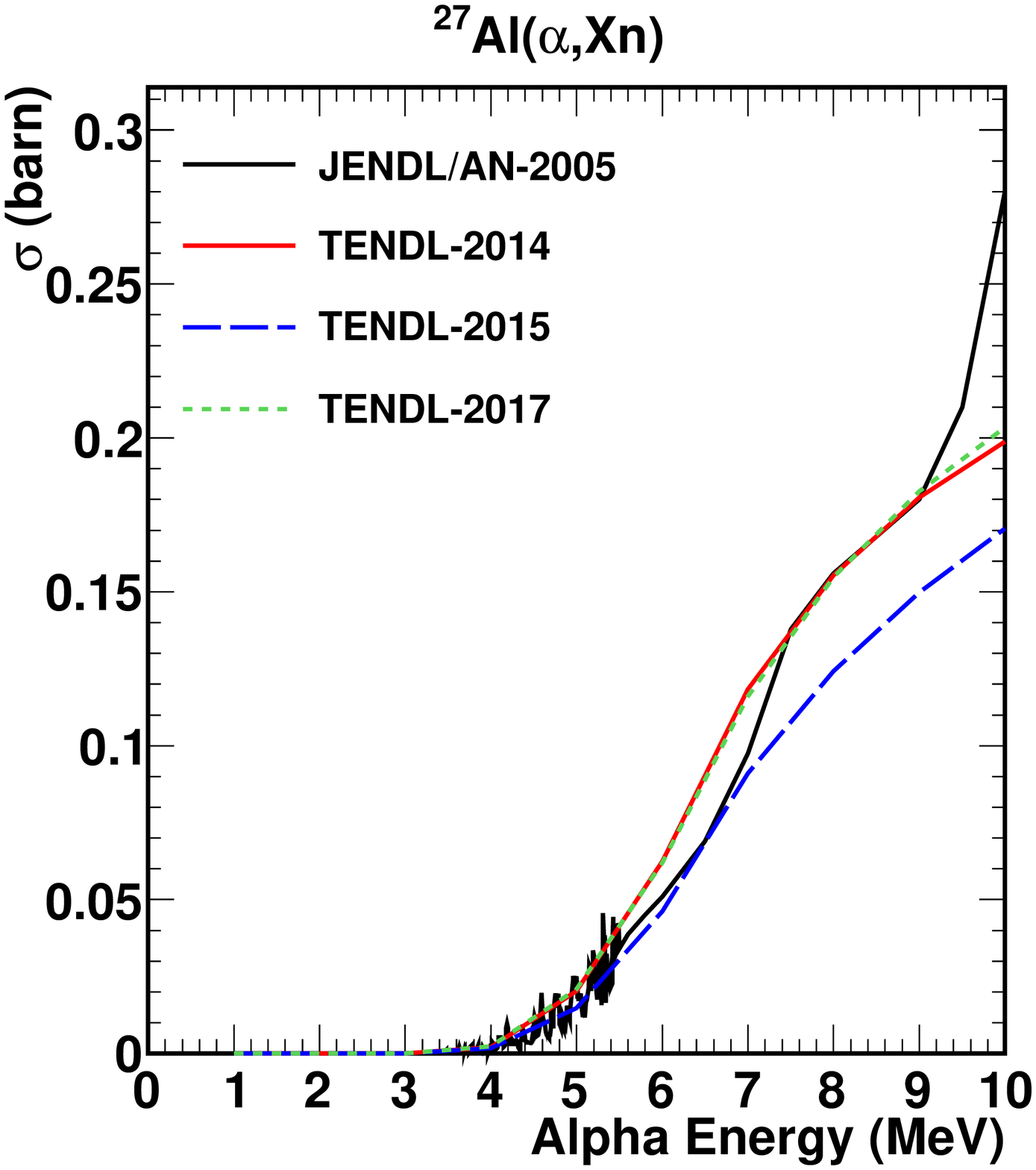} &
\includegraphics[width=0.33\linewidth]{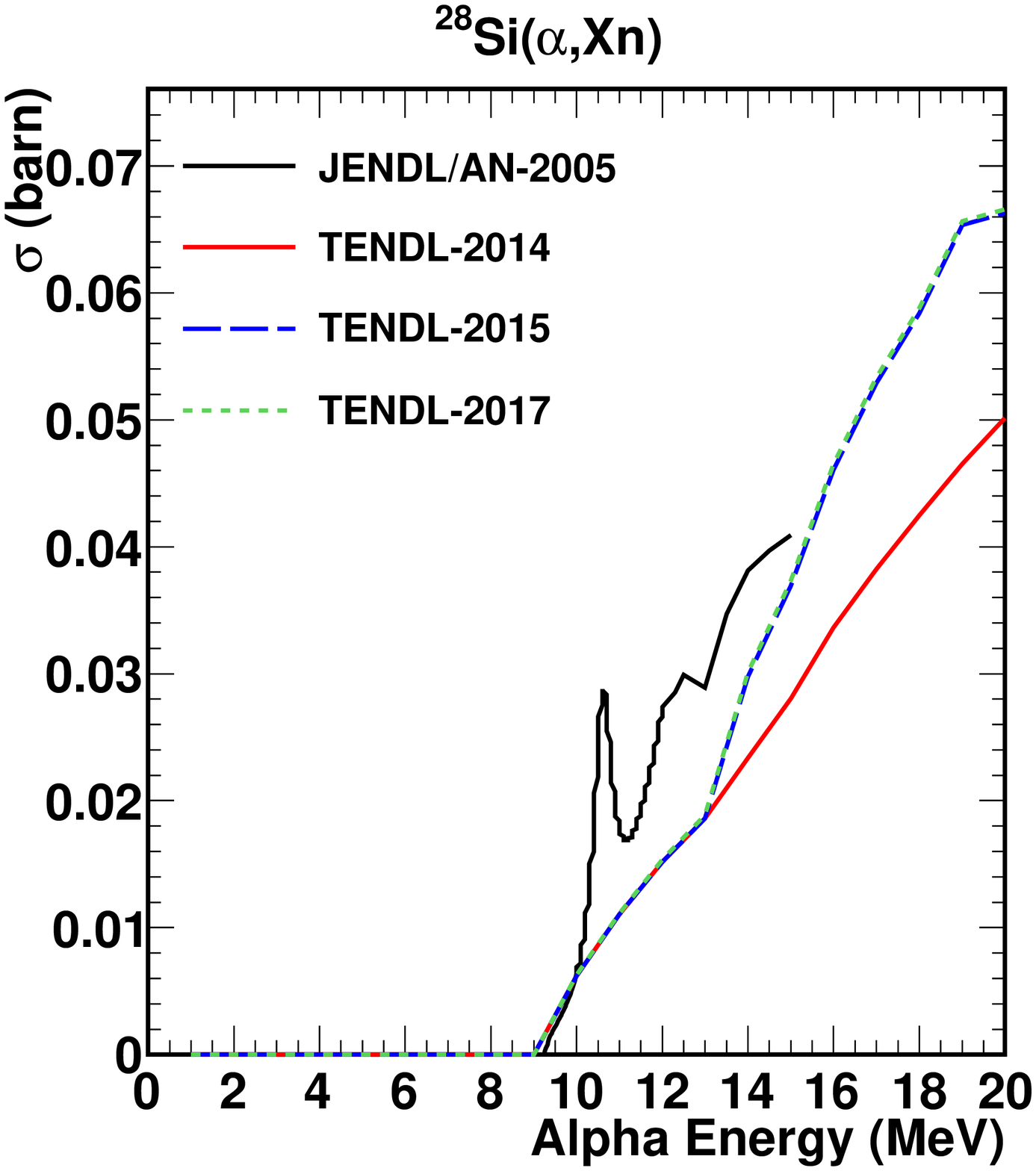} \\
\includegraphics[width=0.33\linewidth]{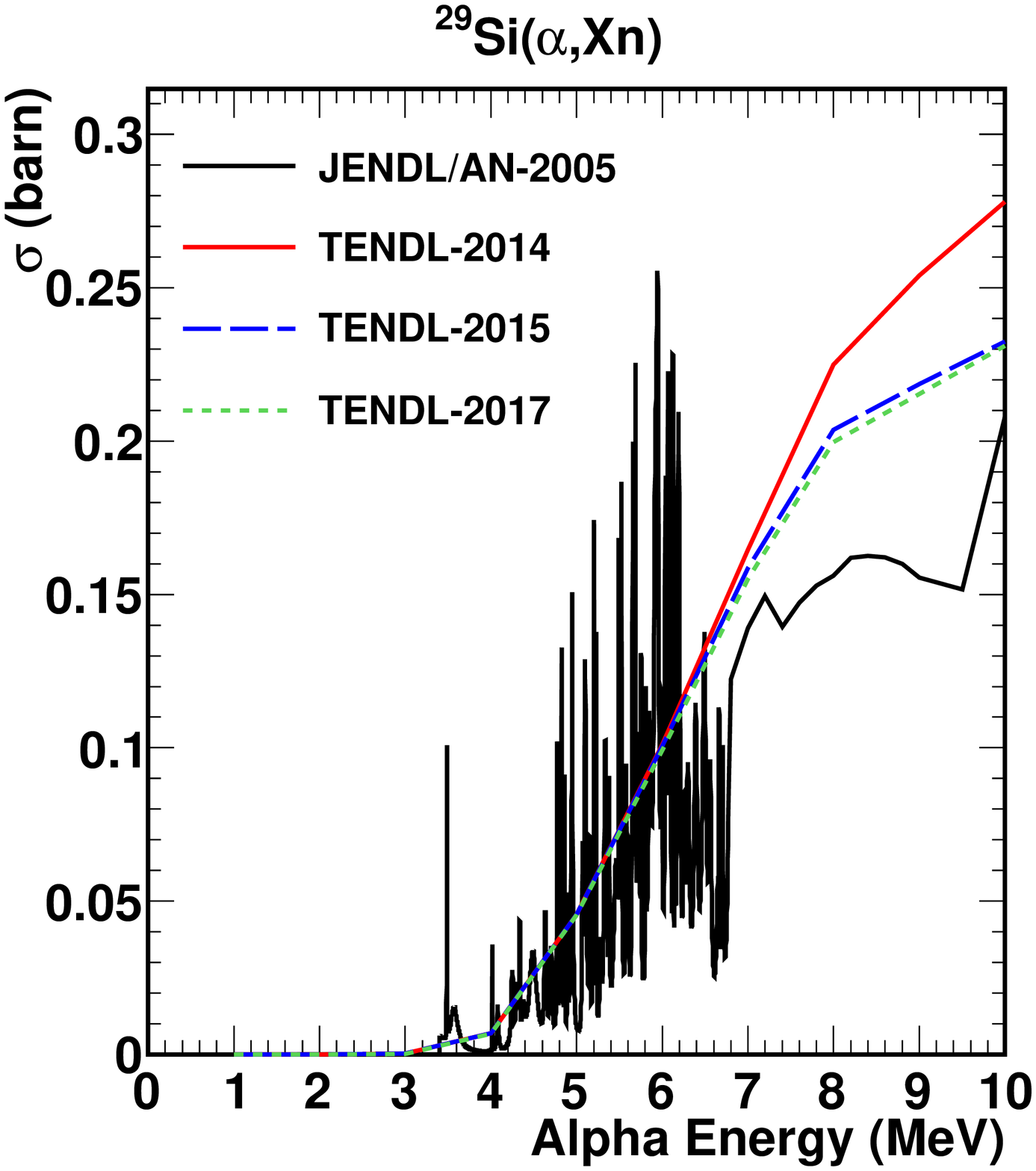} &
\includegraphics[width=0.33\linewidth]{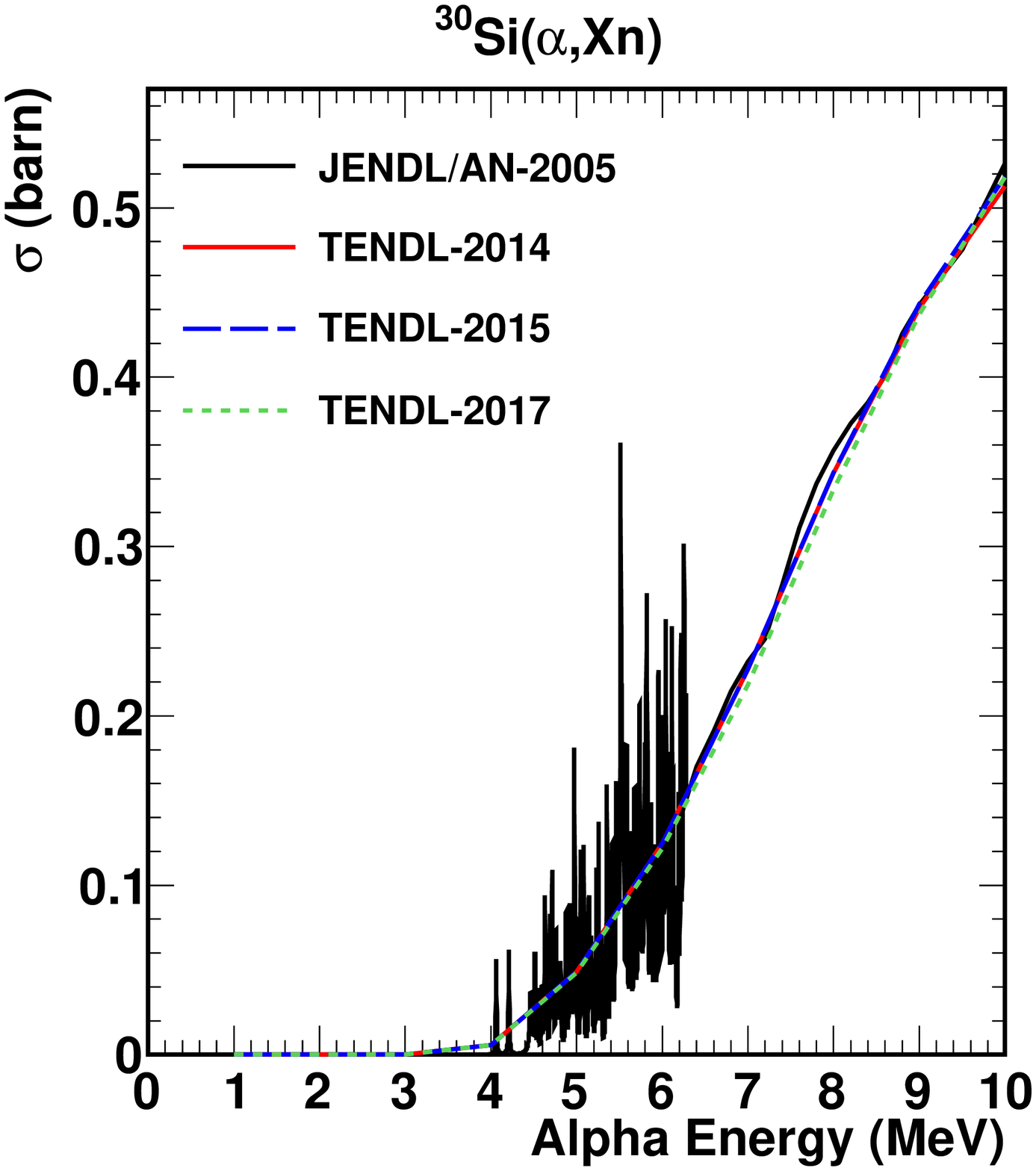} &
\end{array}$
\end{center}
\caption{Neutron production cross sections at low $\alpha$ particle energies for different isotopes and libraries.}
\label{fig:XS_02}
\end{figure*}

\begin{figure*}[htb]
\begin{center}$
\begin{array}{c c c}
\includegraphics[width=0.47\linewidth]{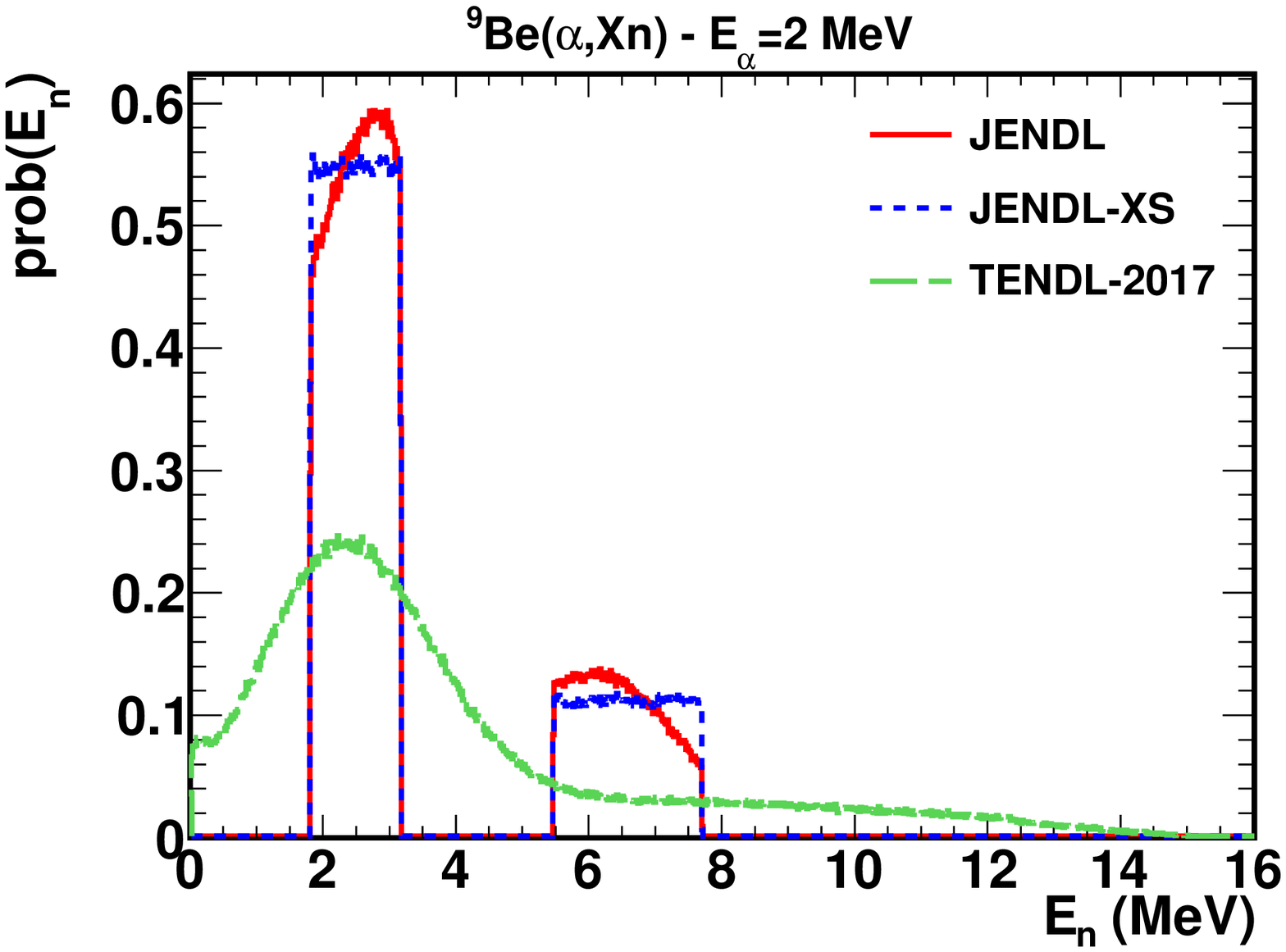} &
\includegraphics[width=0.47\linewidth]{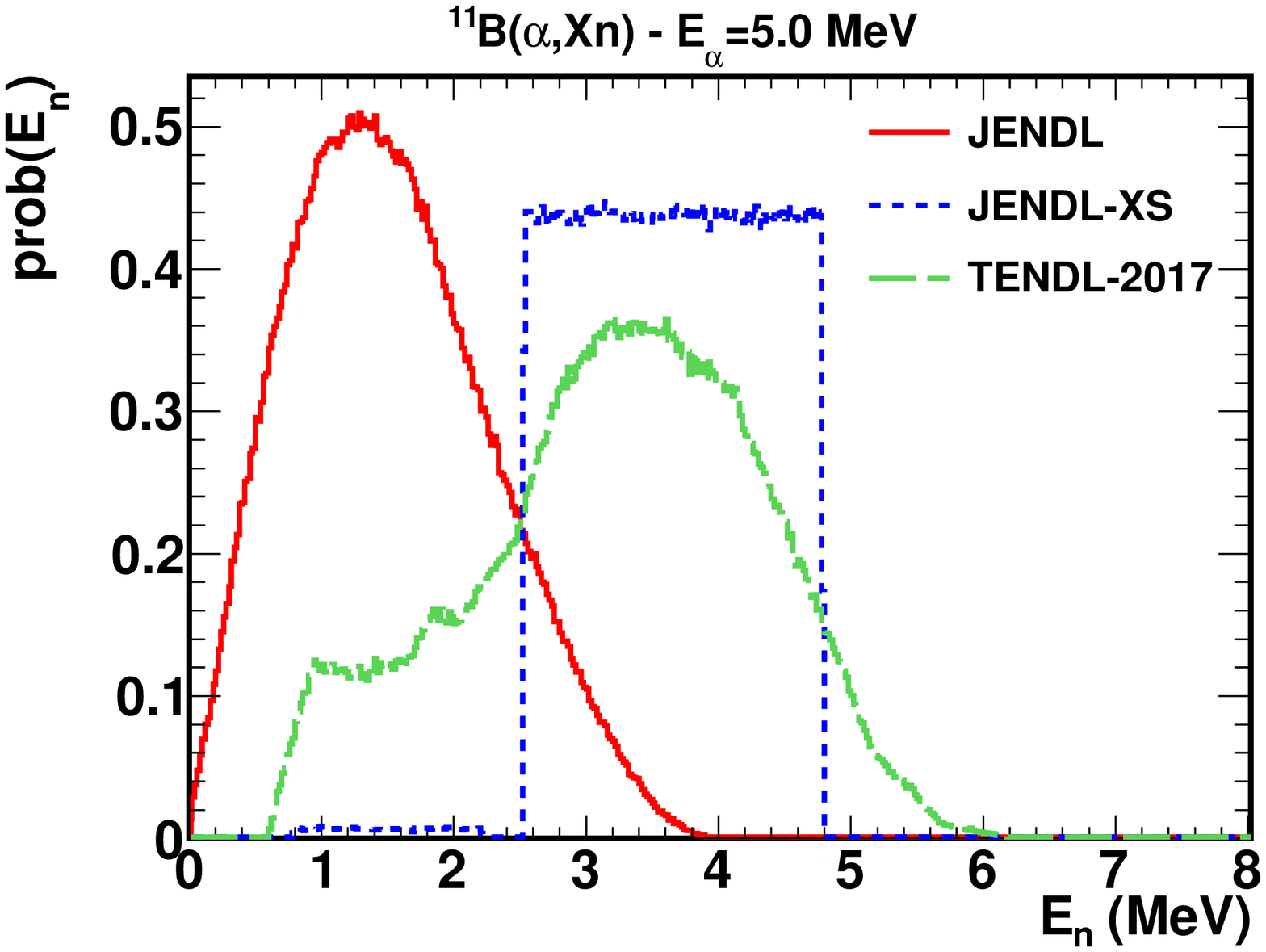} \\
\includegraphics[width=0.47\linewidth]{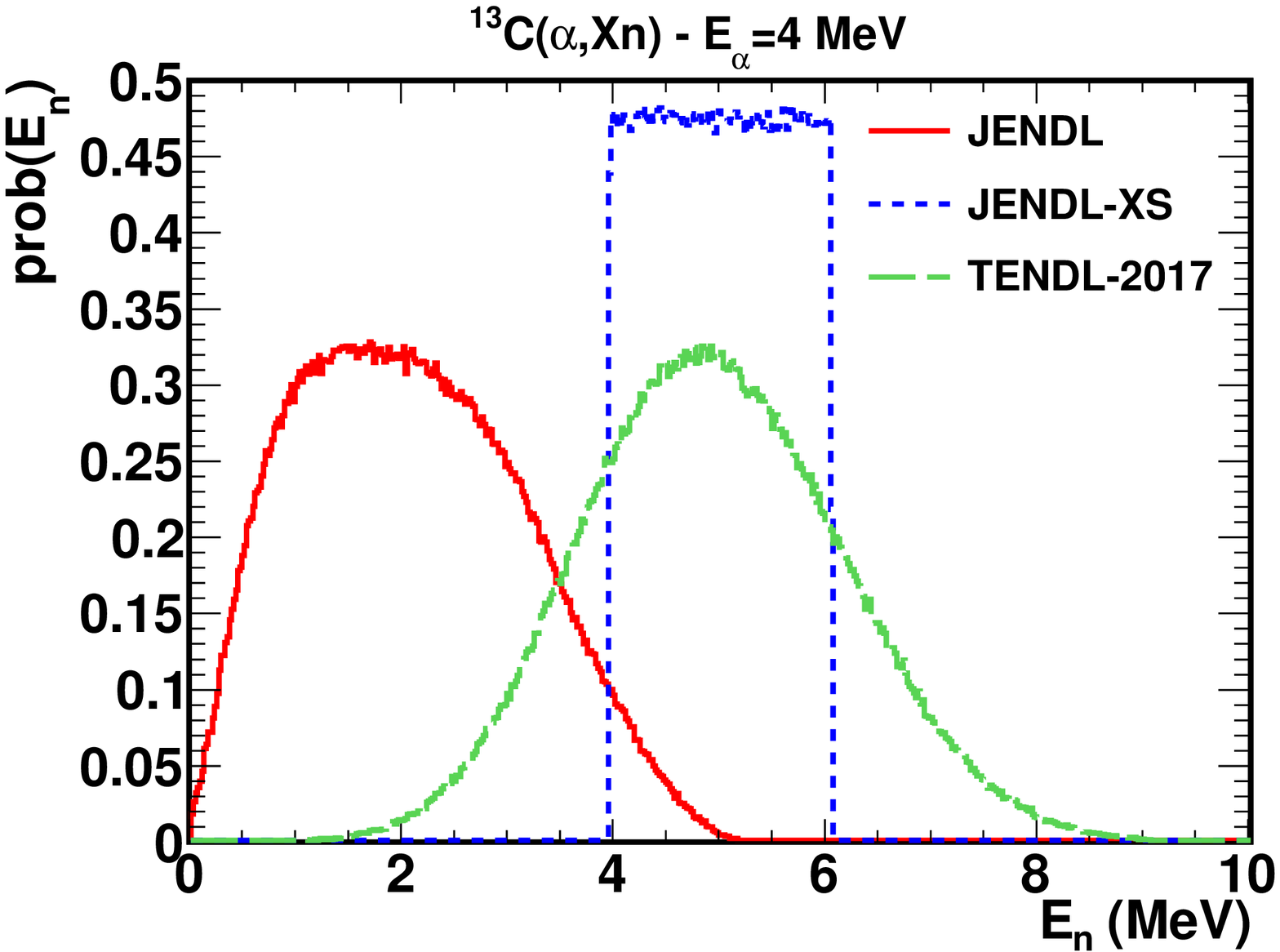} &
\includegraphics[width=0.47\linewidth]{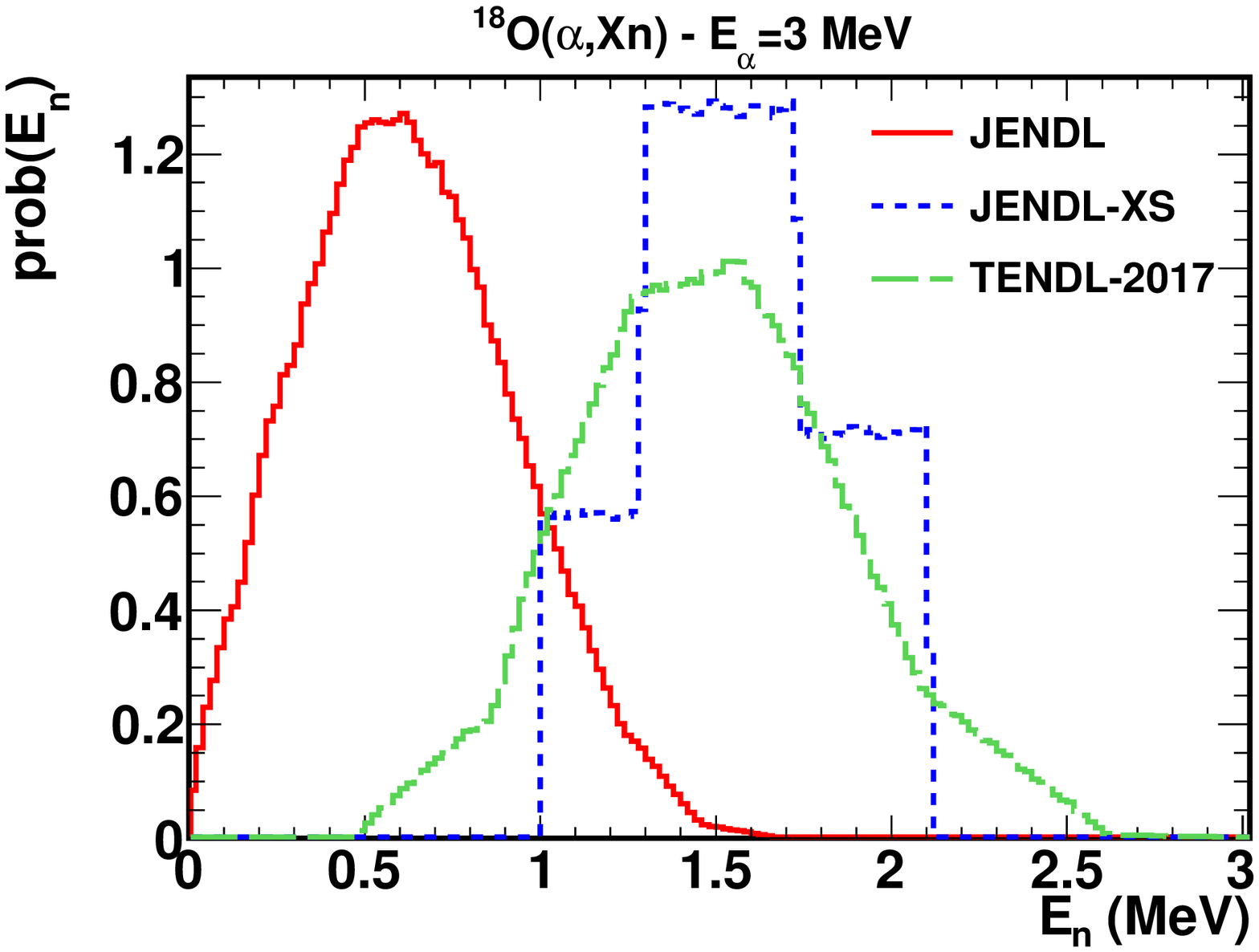} \\
\end{array}$
\end{center}
\caption{Neutron energy spectra induced by ($\alpha$,n) reactions on $^{9}$Be, $^{11}$B, $^{13}$C and $^{18}$O at different $\alpha$-incident fixed energies. The spectra labeled as JENDL correspond to the ones present in the JENDL/AN-2005 library. JENDL-XS denotes the spectra resulting from using the JENDL/AN-2005 partial ($\alpha$,n$_{i}$) cross sections and isotropic neutron angular distribution in the center-of-mass system. The label TENDL-2017 refers to the spectra from the TENDL-2017 library.}
\label{fig:NSpectraLibs}
\end{figure*}

The TENDL and JENDL/AN-2005 neutron production cross sections of all the isotopes included in the JENDL/AN-2005 library are presented in figures~\ref{fig:XS_01} and~\ref{fig:XS_02}. Cross sections are given at low energies, up to 10 MeV in all the cases but $^{12}$C, $^{15}$N and $^{28}$Si, which have larger neutron production thresholds. The cross sections provided by the three TENDL libraries are quite similar for these isotopes, specially the ones corresponding to TENDL-2014 and TENDL-2017. Concerning the comparison between JENDL and TENDL, the latter provides larger neutron production cross sections in most of the cases, with a few exceptions.

It is worth highlighting the differences between the cross sections provided for $^{15}$N. The values from TENDL are two orders of magnitude larger than the values from JENDL/AN-2005. The impact of this discrepancy on the calculation of neutron yields from $\alpha$-decay is not big, since the $^{15}$N($\alpha$,Xn) reaction has a threshold larger than most of the $\alpha$-decay energies. However, these differences show that theoretical calculations could be quite far from reality, and thus neutron yield calculations from ($\alpha$,Xn) reactions in isotopes with no experimental data should be taken with care.

We have also investigated the neutron energy spectra present in the different libraries. In the wide variety of spectra we studied, we observe that JENDL/AN-2005 seems to underestimate the energy of secondary neutrons in some cases. The energy spectra displayed in figure~\ref{fig:NSpectraLibs} are obtained for four distinct nuclei and different energies of the incident $\alpha$ particles. In the four cases the only opened channels are the $^{A}_{Z}$X($\alpha$,n$_{i}$)$^{A+3}_{Z+2}$Y, where $i$ refers to the excited state of the residual nucleus: $i=0$ for the ground state, $i=1$ for the first excited state, and so on. In the JENDL/AN-2005 library each of the ($\alpha$,n$_{i}$) channels appears explicitly, whereas in TENDL-2017 all of them are summed in a single channel. Since there are only two outgoing particles, the energy of a particle emitted with a specific scattering angle is determined by kinematics. In addition to JENDL/AN-2005 and TENDL-2017, we generated the spectra with a third different procedure (JENDL-XS), consisting in using the JENDL/AN-2005 cross sections but not the secondary particle production data. Instead, the neutrons are emitted in Geant4 by assuming an isotropic angular distribution in the center-of-mass system.

For $^{9}$Be (top-left panel) neutrons in JENDL/AN-2005 cover the same energy range as in JENDL-XS, as expected. There are differences in the spectra due the fact that JENDL/AN-2005 accounts for anisotropic angular distribution of neutrons. However, in the other three cases, $^{11}$Be, $^{13}$C and $^{18}$O, these energy ranges differ significantly. The energies of the neutrons from JENDL/AN-2005 are lower than expected from two-body kinematics. On the other hand, the neutron energy spectra from TENDL-2017 have been probably averaged over a range of $\alpha$ particle energies, but the neutron energies are in the expected range. The energy spectra of TENDL-2014 and TENDL-2015 are not very different from those extracted from TENDL-2017.

\FloatBarrier

\section{Comparison with other codes and validation with experimental data \label{Sec:Comparison}}

We have performed several calculations to compare Geant4 with the codes mentioned in Section~\ref{Sec:intro}: NeuCBOT, SOURCES, NEDIS and USD; as well as with experimental data. Since the neutron yields and energy spectra obtained with a code depend mainly on the databases used, we start by describing these databases.

SOURCES and NEDIS calculate the stopping powers using the data from ~\cite{Ziegler1977}, USD and Geant4 from the ICRU 49 report~\cite{ICRU49}, and NeuCBOT from SRIM~\cite{SRIM}. 

\begin{figure}[htb]
\begin{center}
\includegraphics[width=0.62\linewidth]{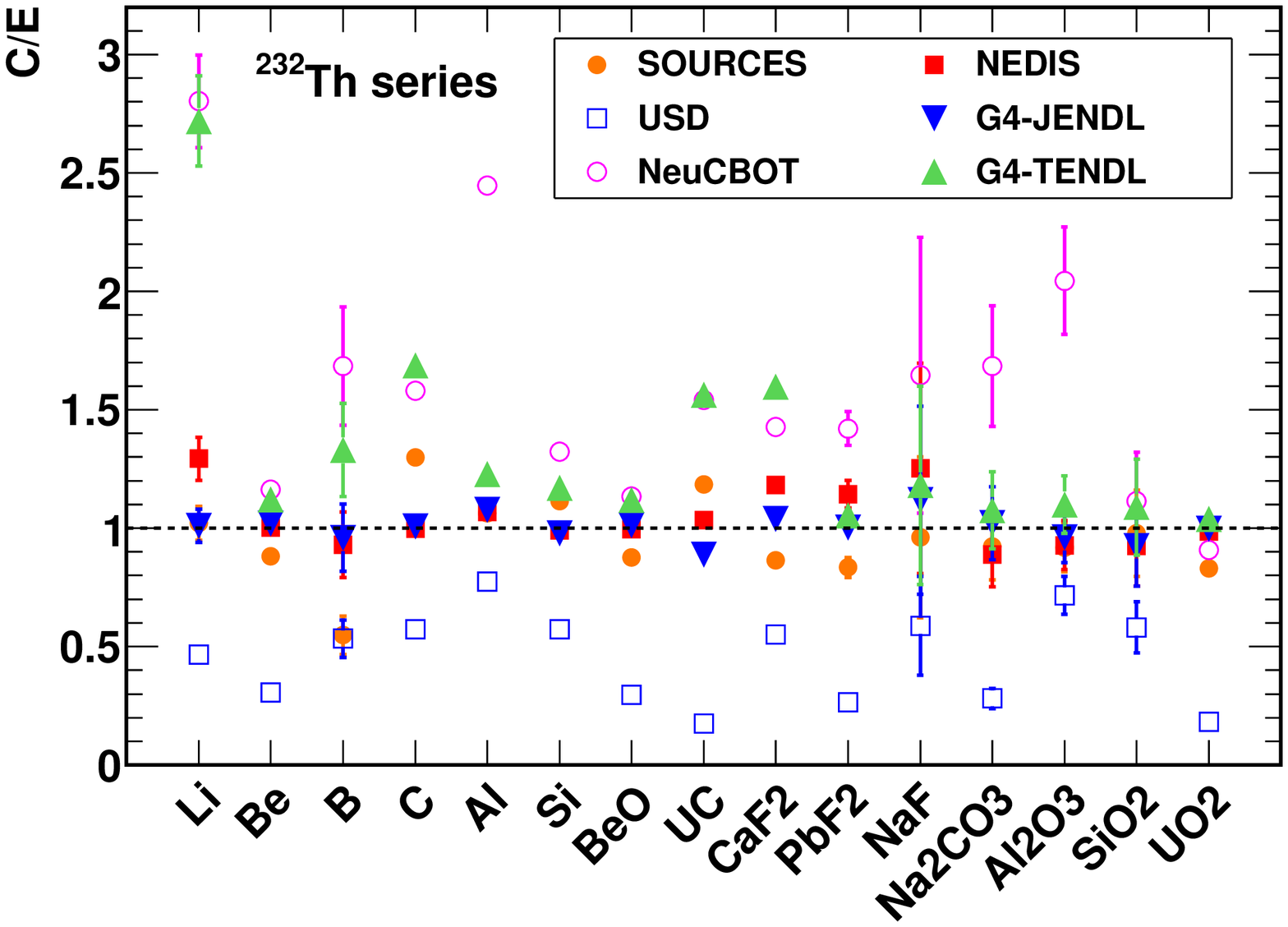} \\
\includegraphics[width=0.62\linewidth]{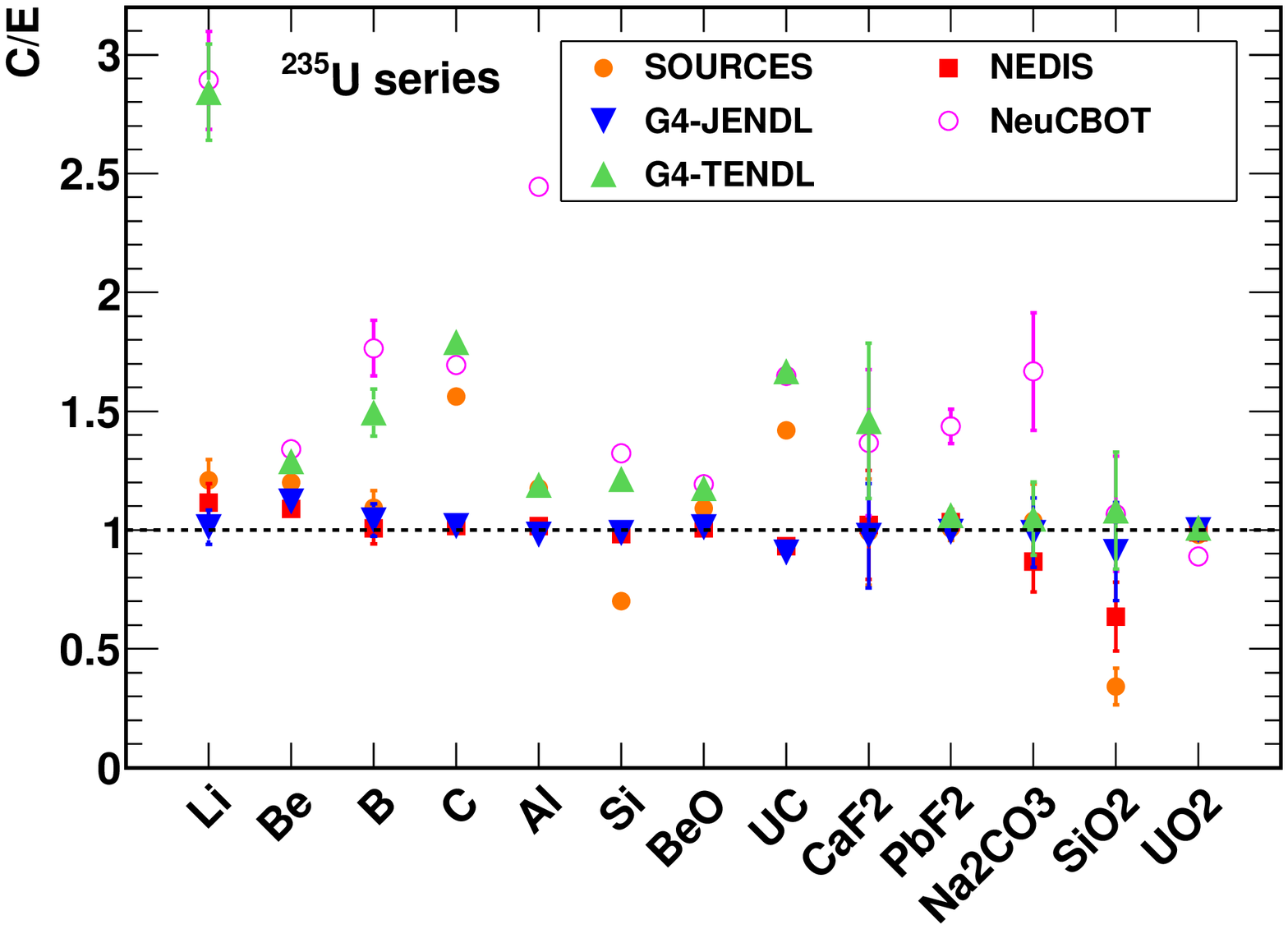} \\
\includegraphics[width=0.62\linewidth]{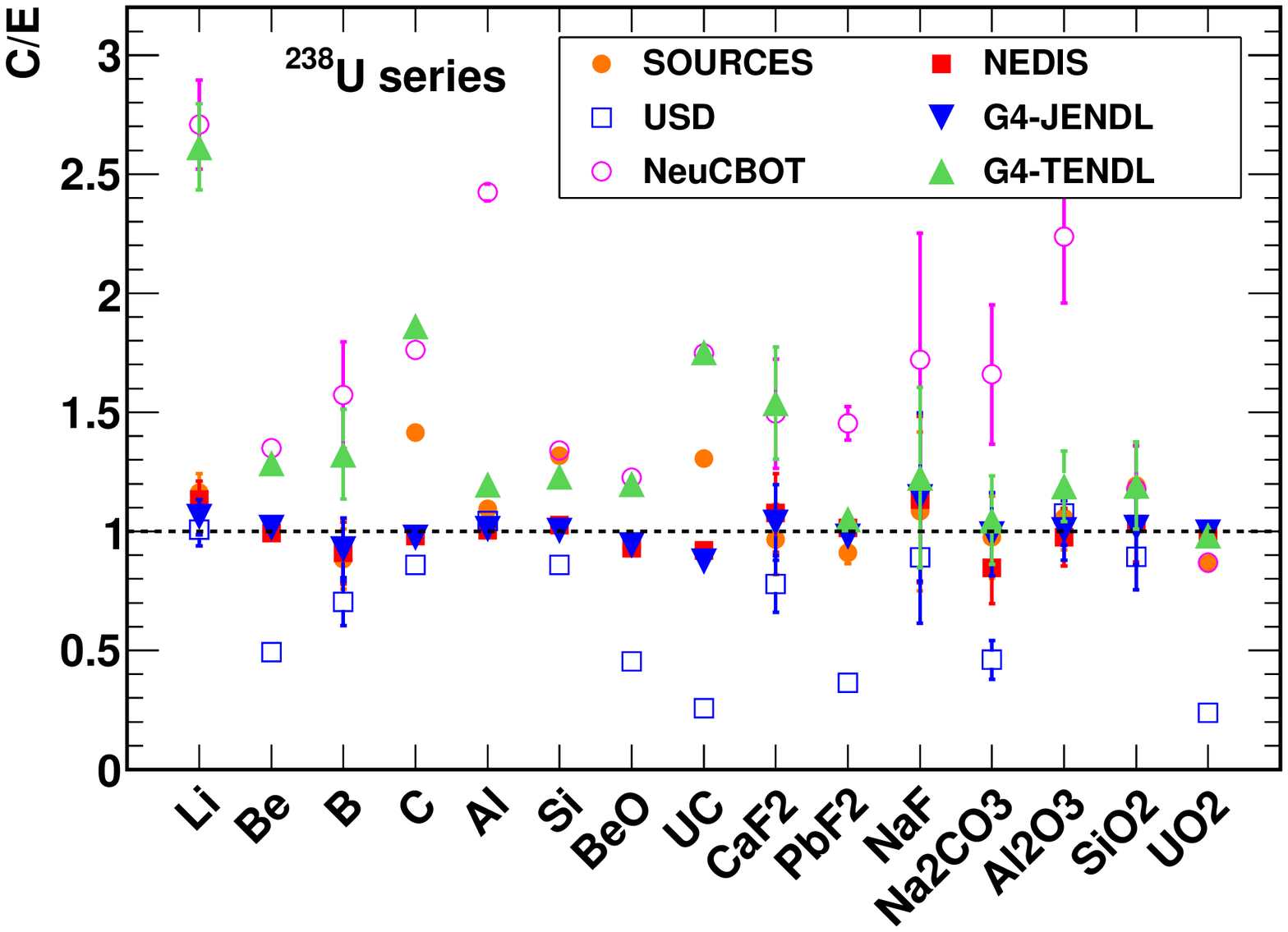}
\end{center}
\caption{Calculated to experimental ratios of the values in Table~\ref{tab:NedisComp_All}. The error bars correspond to the propagation of the uncertainties in the experimental data.}
\label{fig:NedisComp_All_gr}
\end{figure}

The neutron production cross sections and energy spectra are extracted by SOURCES and NEDIS from their own cross section databases~\cite{SOURCES-4C,NEDIS}.  In both cases, the neutron energy spectra are determined by two-body kinematics, from the neutron emission angle. SOURCES assumes an isotropic neutron angular distribution in the center-of-mass system and NEDIS accounts for anisotropic angular distributions in some cases. The cross section database distributed with SOURCES has a limit of 6.5 MeV in the energy of the incident $\alpha$ particle, for most of the isotopes. Hence, they do not cover the energy ranges of the $^{235}$U, $^{238}$U and $^{232}$Th decay series, with $\alpha$ energies up to 7.39, 7.69 and 8.78 MeV, respectively. However, this database has been extended or replaced in various works~\cite{NeuCBOT,Carson_AP,Tomasello_NIMA,Comp_COOLEY,Comp_Fernandes}. Finally, NeuCBOT and USD take the neutron production cross sections and energy spectra from TALYS, whose output strongly depends on the input parameters used. As a consequence, two codes using TALYS do not produce the same result when they use different inputs to generate their databases.

Various comparisons between the different codes and with experimental data are found in~\cite{NeuCBOT,Comp_COOLEY,Comp_Fernandes,Comp_Vlaskin}. In the following lines, we describe the results obtained with Geant4, that extend the comparisons performed in~\cite{Comp_Fernandes} and~\cite{NeuCBOT}.

Fernandes~\textit{et al.} compare in~\cite{Comp_Fernandes} several experimental ($\alpha$,n) yields with calculations performed with SOURCES, NEDIS-2.0 and USD. With SOURCES, they use a modified version of the database~\cite{Carson_AP}, which rises the 6.5 MeV limit up to 10 MeV and utilises new cross sections. We extend their comparison with results from NeuCBOT and Geant4 in Table~\ref{tab:NedisComp_All}. The results correspond to ($\alpha$,n) yields in different materials, for the emission spectra of the $^{232}$Th, $^{235}$U and $^{238}$U decay series in secular equilibrium. In Geant4, we include results from the JENDL/AN-2005 and the TENDL-2017 libraries. The graphs in figure~\ref{fig:NedisComp_All_gr}, depicted using the calculated-to-experimental ratios of the values in Table~\ref{tab:NedisComp_All}, provide a clear overview of our resutls.

SOURCES agrees within 20\% with the data measured for most materials. The largest discrepancies are found in C, UC, Si, SiO$_{2}$ and B, the latter only in the $^{232}$Th series, the one with the largest $\alpha$ particle energies. These discrepancies are due to the C, Si and B neutron production cross sections~\cite{Comp_Fernandes}. NEDIS and G4-JENDL (Geant4 using the JENDL/AN-2005 library) are the results which better reproduce the experimental values, with an agreement better than 10\% in most cases. The only discrepancy larger than 20\% is the one from NEDIS-2.0 in Li in the $^{232}$Th series. USD underestimates the neutron yields in almost all the tested cases, being the discrepancies as large as 80\%. On the contrary, NeuCBOT and G4-TENDL (Geant4 using the TENDL-2017 library) overestimate the neutron yields in most cases. On average, NeuCBOT and G4-TENDL yields are, respectively, 60\% and 40\% larger than the measured values. These discrepancies might be justified by the fact that all threee codes, USD, NeuCBOT and G4-TENDL, use cross sections from TALYS, but presumably with different input parameters.

\begin{table*}[ht]
\resizebox{\linewidth}{!}{%
\centering\
\begin{tabular}{ccccccccc}
\hline
\multirow{2}{*}{Target} & Decay & \multirow{2}{*}{Measured} & \multirow{2}{*}{SOURCES} & \multirow{2}{*}{NEDIS} & \multirow{2}{*}{USD}  & \multirow{2}{*}{NeuCBOT} & Geant4 & Geant4 \\ 
                        &      Chain                     &                           &                          &                        &                       &                          & JENDL  & TENDL-2017 \\
\hline
\multirow{3}{*}{Li} & $^{232}$Th & 11.99(84) & 12.25 & 15.5 & 5.60 & 33.6 & 12.1 & 32.6 \\
 & $^{235}$U & 9.82(70) & 11.88 & 10.95 & & 28.4 & 9.93 & 27.9 \\
 & $^{238}$U & 6.35(44) & 7.375 & 7.19 & 6.40 & 17.2 & 6.73 & 16.6 \\
\hline
\multirow{3}{*}{Be} & $^{232}$Th & 117.7(20) & 103.7 & 118 & 35.9 & 137 & 120 & 132  \\
 & $^{235}$U & 100.0(17) & 120.0 & 109 & & 134 & 112 & 129  \\
 & $^{238}$U & 80.1(13) & 79.93 & 79.6 & 39.5 & 108 & 81.2 & 103  \\
\hline
\multirow{3}{*}{B} & $^{232}$Th & 27(4) & 14.81 & 25.1 & 14.4  & 45.5 & 25.9 & 35.9    \\
 & $^{235}$U & 24.3(16) & 26.61 & 24.5 & & 42.9 & 25.3 & 36.3   \\
 & $^{238}$U & 21(3) & 18.57 & 19.1 & 14.8 & 33.0 & 19.4 & 27.8    \\
\hline
\multirow{3}{*}{C} & $^{232}$Th & 0.2305(32) & 0.2991 & 0.23 & 0.132 & 0.364 & 0.232 & 0.389    \\
 & $^{235}$U & 0.1965(28) & 0.3070 & 0.20 & & 0.333 & 0.200 & 0.352    \\
 & $^{238}$U & 0.1397(20) & 0.1977 & 0.137 & 0.120 & 0.246 & 0.136 & 0.260    \\
\hline
\multirow{3}{*}{Al} & $^{232}$Th & 3.086(14) & 3.282 & 3.3 & 2.39 & 7.55 & 3.33 & 3.79    \\
 & $^{235}$U & 2.409(28) & 2.832 & 2.45 & & 5.89 & 2.36 & 2.87    \\
 & $^{238}$U & 1.514(22) & 1.657 & 1.52 & 1.58 & 3.67 & 1.53 & 1.81   \\
\hline
\multirow{3}{*}{Si} & $^{232}$Th & 0.4336(52) & 0.4830 & 0.43 & 0.248 & 0.574 & 0.424 & 0.507    \\
 & $^{235}$U & 0.3258(39) & 0.2286 & 0.32 & & 0.431 & 0.322 & 0.396    \\
 & $^{238}$U & 0.2047(25) & 0.2696 & 0.21 & 0.176 & 0.274 & 0.205 & 0.252    \\
\hline
\multirow{3}{*}{BeO} & $^{232}$Th & 43.40(64) & 38.03 & 43.2 & 12.9 & 49.2 & 43.9 & 48.6    \\
 & $^{235}$U & 40.42(61) & 44.10 & 40.7 & & 48.2 & 40.9 & 47.5   \\
 & $^{238}$U & 31.56(30) & 29.37 & 29.3 & 14.3 & 38.7 & 29.7 & 37.8    \\
\hline
\multirow{3}{*}{UC} & $^{232}$Th & 0.03665(55) & 0.04339 & 0.0379 & 0.00638 & 0.0565 & 0.0326 & 0.0573    \\
 & $^{235}$U & 0.03168(48) & 0.04498 & 0.0295 & & 0.0522 & 0.0287 & 0.0528   \\
 & $^{238}$U & 0.02231(33) & 0.02913 & 0.0205 & 0.00576 & 0.0390 & 0.0195 & 0.0391    \\
\hline
\multirow{3}{*}{CaF$_{2}$} & $^{232}$Th & 10.58(18) & 9.136 & 12.5 & 5.82 & 15.1 & 11.0 & 16.9    \\
 & $^{235}$U & 9.8(22) & 9.719 & 10.0 & & 13.4 & 9.57 & 14.3    \\
 & $^{238}$U & 6.01(92) & 5.802 & 6.48 & 4.68 & 8.98 & 6.23 & 9.25    \\
\hline
\multirow{3}{*}{PbF$_{2}$} & $^{232}$Th & 6.97(35) & 5.813 & 7.97 & 1.85 & 9.90 & 6.98 & 7.35    \\
 & $^{235}$U & 6.17(31) & 6.207 & 6.38 & & 8.86 & 6.13 & 6.56    \\
 & $^{238}$U & 4.08(20) & 3.710 & 4.14 & 1.49 & 5.93 & 3.99 & 4.29   \\
\hline
\multirow{2}{*}{NaF} & $^{232}$Th & 12.7(45) & 12.21 & 15.9 & 7.47 & 20.9 & 14.2 & 15.0    \\
 & $^{238}$U & 6.8(21) & 7.369 & 7.7 & 6.05 & 11.7 & 7.78 & 8.33    \\
\hline
\multirow{3}{*}{Na$_{2}$CO$_{3}$} & $^{232}$Th & 3.3(5) & 3.043 & 2.93 & 0.928 & 5.56 & 3.37 & 3.55    \\
 & $^{235}$U & 2.7(4) & 2.805 & 2.34 & & 4.50 & 2.67 & 2.83   \\
 & $^{238}$U  & 1.7(3) & 1.660 & 1.44 & 0.783 & 2.82 & 1.68 & 1.78    \\
\hline
\multirow{2}{*}{Al$_{2}$O$_{3}$} & $^{232}$Th & 1.8(2) & 1.651 & 1.67 & 1.29 & 3.68 & 1.73 & 1.98    \\
 & $^{238}$U & 0.8(1) & 0.8437 & 0.78 & 0.860 & 1.79 & 0.803 & 0.951    \\
\hline
\multirow{3}{*}{SiO$_{2}$} & $^{232}$Th & 0.27(5) & 0.2641 & 0.25 & 0.157 & 0.301 & 0.250 & 0.294    \\
 & $^{235}$U & 0.22(5) & 0.07523 & 0.14 & & 0.235 & 0.200 & 0.238    \\
 & $^{238}$U & 0.13(2) & 0.1550 & 0.134 & 0.116 & 0.153 & 0.132 & 0.155    \\
\hline
\multirow{3}{*}{UO$_{2}$} & $^{232}$Th & 0.03446(41) & 0.02865 & 0.034 & 0.00630 & 0.0313 & 0.0344 & 0.0358    \\
 & $^{235}$U & 0.03230(22) & 0.03169 & 0.032 & & 0.0287 & 0.0323 & 0.0326    \\
 & $^{238}$U & 0.02312(28) & 0.01997 & 0.0228 & 0.00548 & 0.0201 & 0.0230 & 0.0227    \\
\hline
\end{tabular}}
\caption{Measured and calculated neutron yields (neutrons per 10$^{6}$ $\alpha$-decays) for the spectra of the $^{232}$Th, $^{235}$U and $^{238}$U decay series in secular equilibrium. The measured values and those of SOURCES, NEDIS and USD are from~\cite{Comp_Fernandes}.}
\label{tab:NedisComp_All}
\end{table*}

\begin{figure*}[htb]
\begin{center}$
\begin{array}{c c c}
\includegraphics[width=0.33\linewidth]{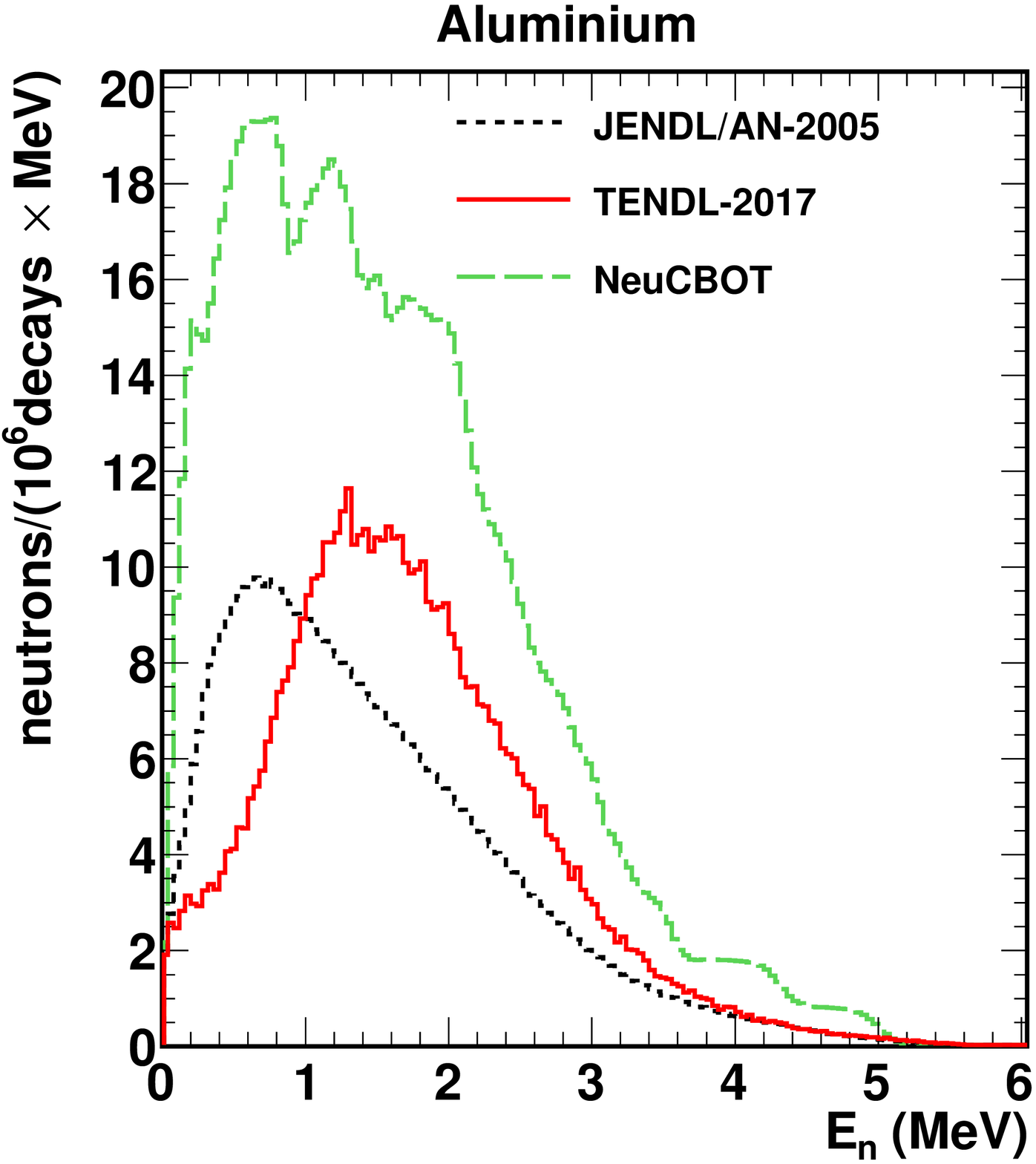} &
\includegraphics[width=0.33\linewidth]{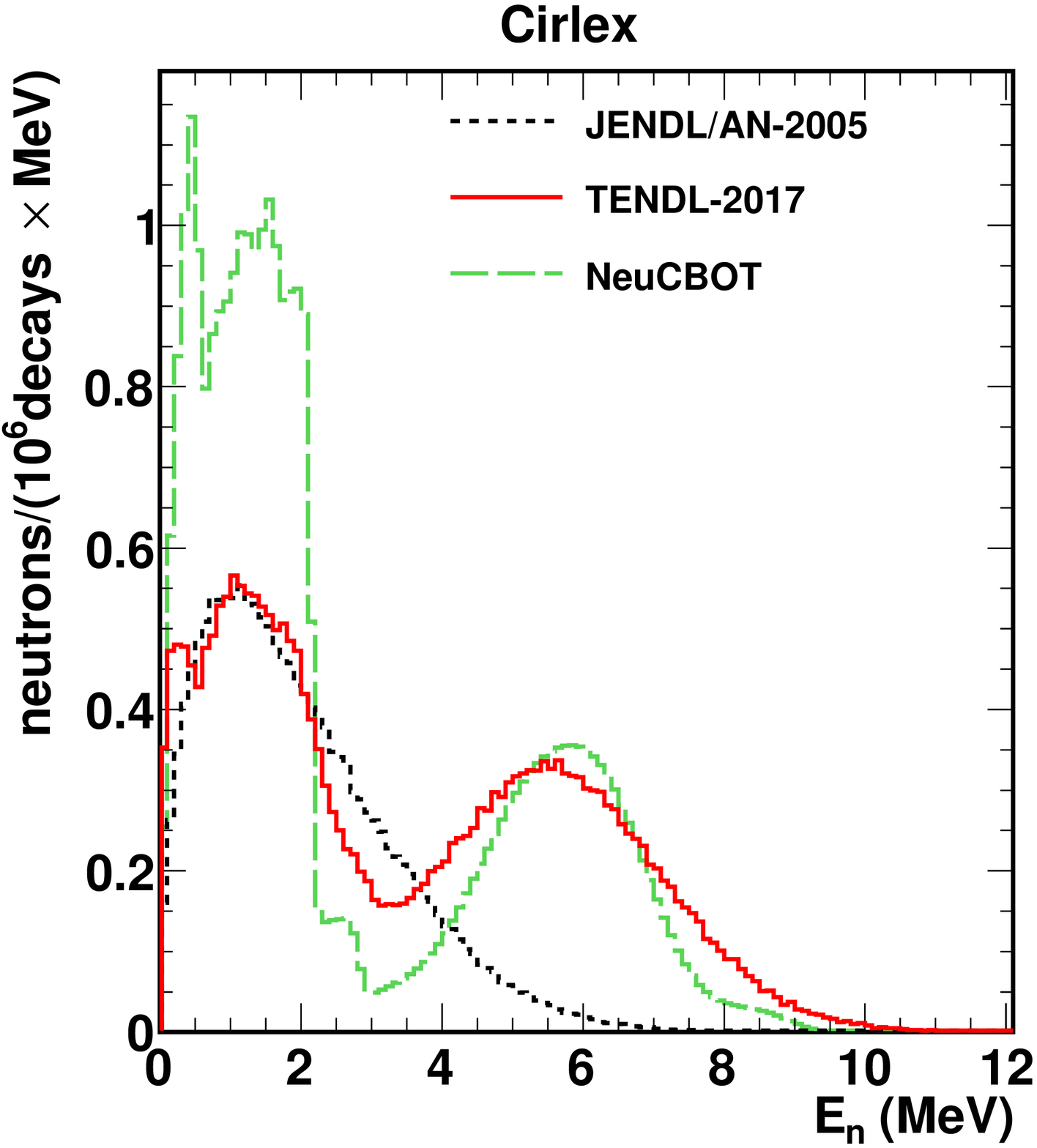} &
\includegraphics[width=0.33\linewidth]{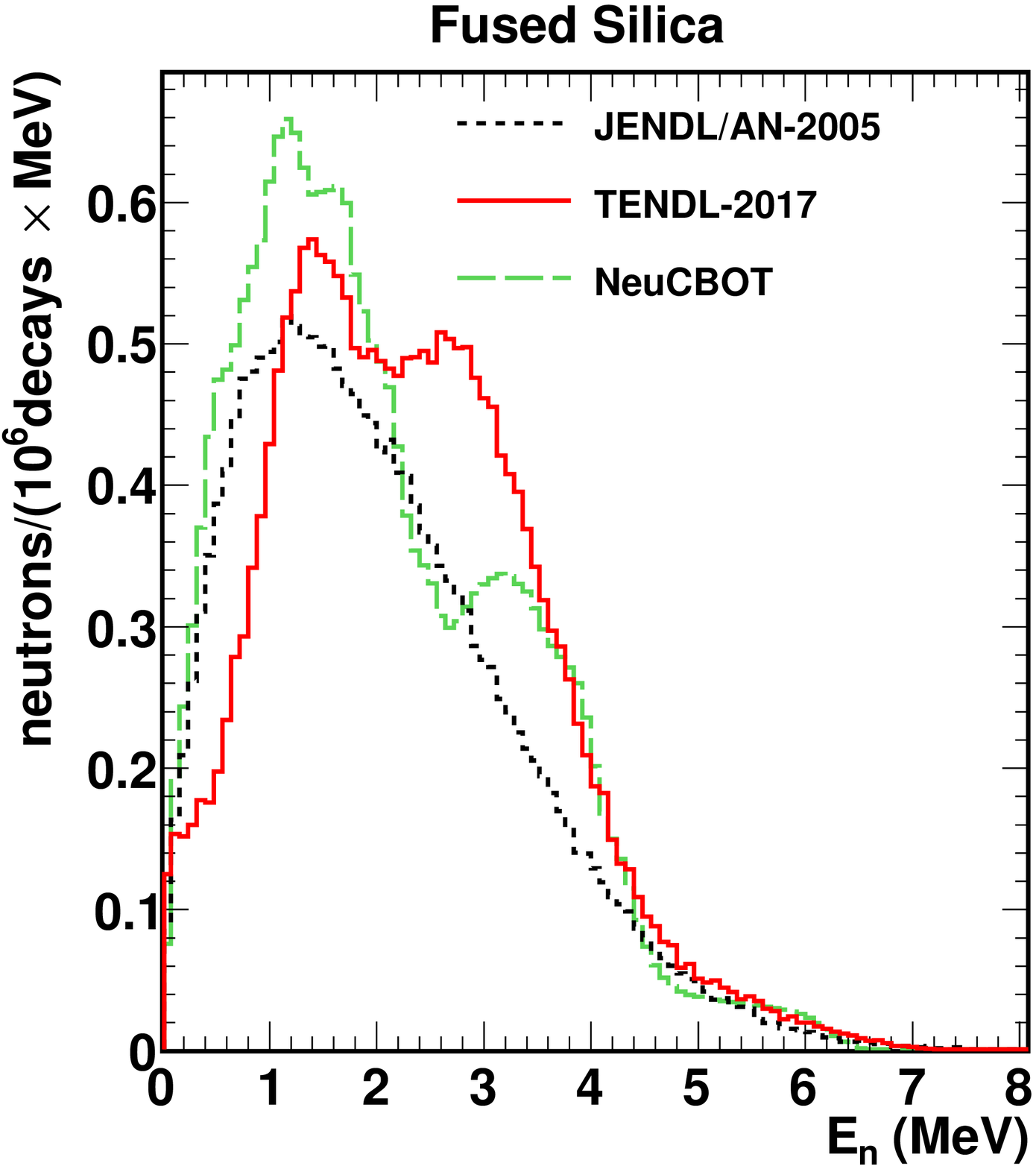} \\
\includegraphics[width=0.33\linewidth]{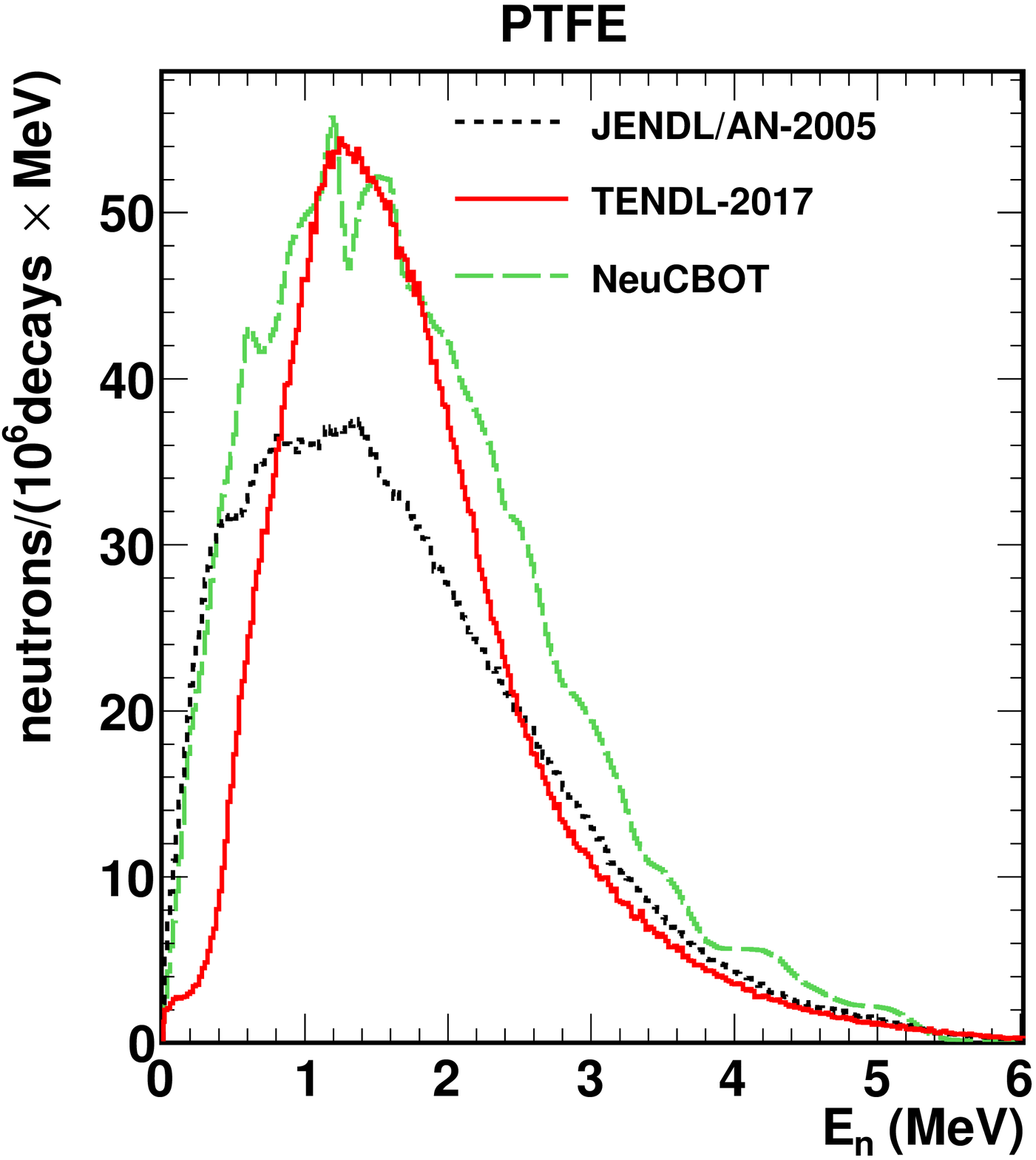} &
\includegraphics[width=0.33\linewidth]{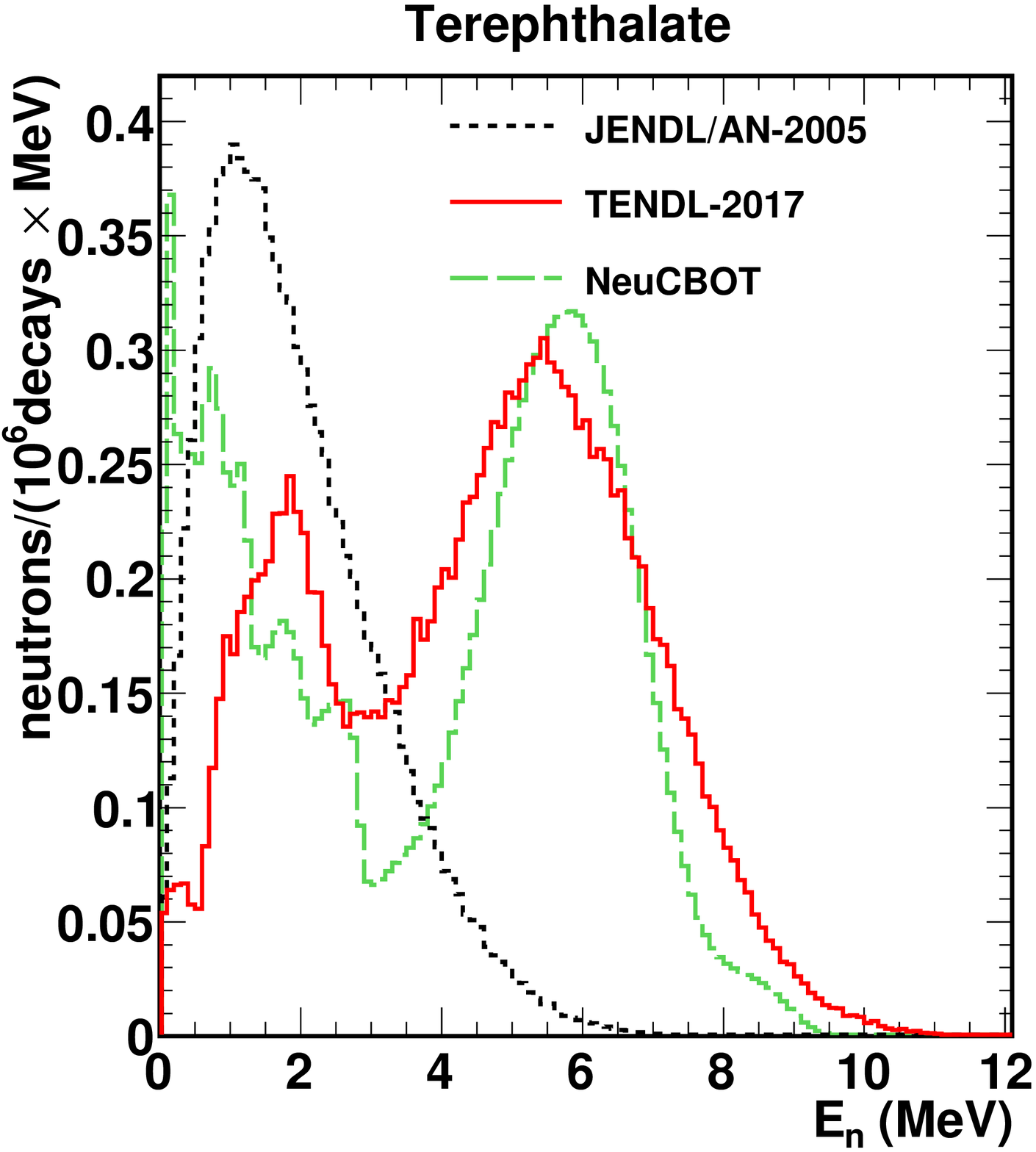} &
\includegraphics[width=0.33\linewidth]{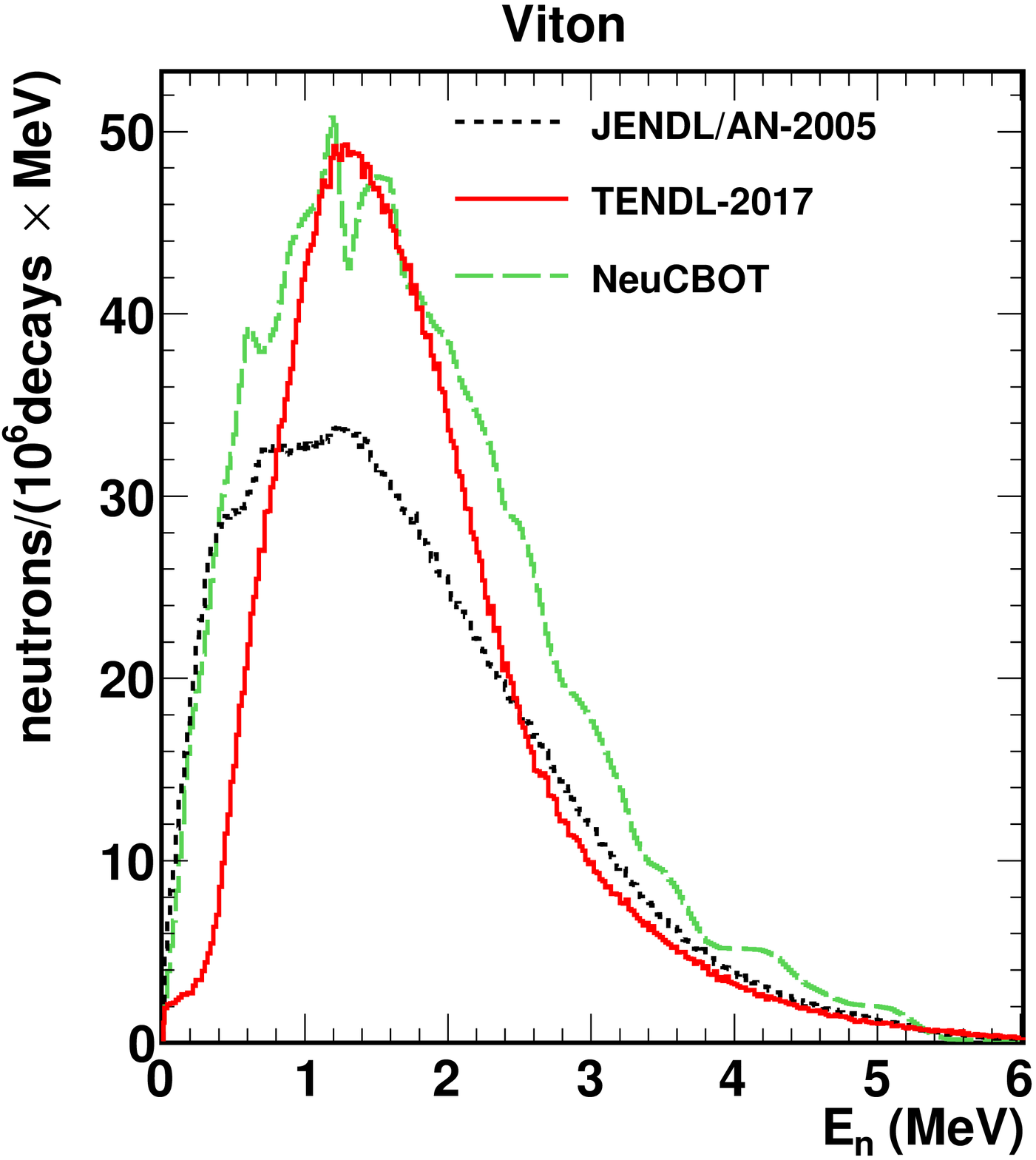} \\
\end{array}$
\end{center}
\caption{Neutron energy spectra induced by ($\alpha$,n) reactions corresponding to the $^{232}$Th series in secular equilibrium on different materials. The spectra have been obtained with NeuCBOT and with Geant4, using in the latter case the JENDL/AN-2005 and TENDL-2017 libraries.}
\label{fig:Comp_WesterdaleSpectra}
\end{figure*}

We also consider the comparison performed by Westerdale and Meyers in~\cite{NeuCBOT}. It focuses in materials common in low-background experiments, and it is performed between NeuCBOT, SOURCES and analytical calculations with experimental data. The version of SOURCES used here extends the neutron production cross section above 6.5 MeV with data from JENDL/AN-2005. They consider the same decay series as in the previous case, $^{232}$Th, $^{235}$U and $^{238}$U, but now the $^{238}$U decay chain is splitted into an upper and a lower chain. The upper chain $^{238}$U$_{upper}$ contains all isotopes before $^{226}$Ra and the lower chain $^{238}$U$_{lower}$, $^{226}$Ra and its progeny. In addition, they also account for the $^{210}$Pb decay chain, which is the last part of $^{238}$U$_{lower}$, consisting of $^{210}$Pb and its progeny. This splitting is performed because in practice secular equilibrium of the $^{238}$U decay chain may be broken due to different processes, but it is expected to be preserved in each sub-chain.

The results are shown in Table~\ref{tab:Westerdale_Comp}. The list of materials is larger in~\cite{NeuCBOT} since only those with data in the JENDL/AN-2005 library are considered. The values in the table correspond to ($\alpha$,n) yields in different materials, per 10$^{6}$ $\alpha$-decays, for the emission spectra of the different decay chains. This normalization is the same as in Table~\ref{tab:NedisComp_All} but different than the one in~\cite{NeuCBOT}, where the yields are given per decay of the parent nucleus, i.e. per decay of $^{232}$Th in the $^{232}$Th series, per decay of $^{235}$U in the $^{235}$U series, and so on. The two normalizations are related by a factor of 6 ($^{232}$Th), 7 ($^{235}$U), 3 ($^{238}$U$_{upper}$), 5 ($^{238}$U$_{lower}$) and 1 ($^{210}$Pb).

We take the yields of NeuCBOT and SOURCES from~\cite{NeuCBOT}, and extend the comparison with Geant4 results. We use the two libraries JENDL/AN-2005 (G4-JENDL) and TENDL-2017 (G4-TENDL-2017), as they are those used in Table~\ref{tab:NedisComp_All}. For some low-background experiments it is important to distinguish between neutrons emitted in coincidence with $\gamma$-rays and neutrons generated without $\gamma$-ray emission~\cite{NeutronVetoExample,DarkSide50}. For this reason, we also calculate the contribution to the G4-JENDL yields of the reactions that produce neutrons without emitting $\gamma$-rays, which in our case is predominantly the ($\alpha$,n$_{0}$) reaction, with small additional contributions of the ($\alpha$,np) or ($\alpha$,n$\alpha$) channels in some cases. This contribution is calculated only when the reaction channels involved appear separately. In JENDL/AN-2005, this occurs in all the nuclei but $^{27}$Al and $^{28,29,30}$Si.

The neutron yields from SOURCES and G4-JENDL are very similar, presumably because both results are based on very similar cross sections. The neutron yields from NeuCBOT are the highest ones in almost all cases, producing from 1.2 to 2.4 more neutrons than G4-JENDL and SOURCES. G4-TENDL-2017 values are in between, around 1.2 to 1.6 times larger than the ones of G4-JENDL and SOURCES, in most cases. These results show the same trend as in Table~\ref{tab:NedisComp_All}, indicating that TENDL-2017 and NeuCBOT overestimate the neutron yields up to factors of 2-3, depending on the material and the decay chain. The calculations performed with JENDL/AN-2005 but taking into account only reactions without $\gamma$-ray emission (G4-JENDL$^{*}$) provide values from 2 to 4 times smaller than those of standard JENDL/AN-2005, and from 3 to 6 times smaller than those of NeuCBOT.

\begin{table*}[ht]
\centering\
\begin{tabular}{llccccc}
\hline
Material & code & $^{232}$Th & $^{235}$U & $^{238}$U$_{upper}$ & $^{238}$U$_{lower}$ & $^{210}$Pb \\
\hline
\multirow{5}{*}{\shortstack{Alumina \\ (Al$_{2}$O$_{3}$)}}
                              & NeuCBOT          &   3.68 & 2.87 & 0.171 & 2.76 & 0.735  \\
                              & SOURCES          &   1.64 & 1.22 & 0.075 & 1.23 & 0.299  \\
                              & G4-TENDL-2017    &   1.98 & 1.47 & 0.094 & 1.46 & 0.370  \\
                              & G4-JENDL         &   1.73 & 1.21 & 0.073 & 1.23 & 0.294  \\
\hline
\multirow{5}{*}{\shortstack{Aluminum \\ (Al)}}
                              & NeuCBOT          &   7.55 & 5.89 & 0.333 & 5.66 & 1.49  \\
                              & SOURCES          &   3.25 & 2.39 & 0.116 & 2.42 & 0.55  \\
                              & G4-TENDL-2017    &   3.79 & 2.87 & 0.156 & 2.80 & 0.70  \\
                              & G4-JENDL         &   3.33 & 2.36 & 0.111 & 2.39 & 0.53  \\
\hline
\multirow{5}{*}{\shortstack{Cirlex \\ (Atom fractions (\%): \\ H:25.4, C:56.6 \\ N:5.1, O:12.9)}}
                              & NeuCBOT          &   0.515 & 0.367 & 0.089 & 0.402 & 0.139  \\
                              & SOURCES          &   0.268 & 0.204 & 0.041 & 0.216 & 0.075  \\
                              & G4-TENDL-2017    &   0.419 & 0.318 & 0.093 & 0.344 & 0.146  \\
                              & G4-JENDL         &   0.270 & 0.195 & 0.036 & 0.208 & 0.068  \\
                              & G4-JENDL$^{*}$   &   0.150 & 0.112 & 0.033 & 0.122 & 0.055  \\
\hline
\multirow{5}{*}{\shortstack{Fused Silica \\ (SiO$_{2}$)}}
                              & NeuCBOT          &   0.302 & 0.234 & 0.025 & 0.230 & 0.079  \\
                              & SOURCES          &   0.245 & 0.206 & 0.029 & 0.199 & 0.074  \\
                              & G4-TENDL-2017    &   0.294 & 0.238 & 0.029 & 0.230 & 0.090  \\
                              & G4-JENDL         &   0.250 & 0.200 & 0.026 & 0.196 & 0.072  \\
\hline
\multirow{5}{*}{\shortstack{PTFE \\ (CF$_{2}$)}}
                              & NeuCBOT          &   21.2 & 18.7 & 3.87 & 17.7 & 9.37  \\
                              & SOURCES          &   14.9 & 13.6 & 2.40 & 12.8 & 6.08  \\
                              & G4-TENDL-2017    &   16.4 & 14.6 & 2.61 & 13.7 & 6.72  \\
                              & G4-JENDL         &   15.5 & 13.6 & 2.41 & 12.7 & 6.04  \\
                              & G4-JENDL$^{*}$   &   3.62 & 3.46 & 1.00 & 3.25 & 2.11  \\
\hline
\multirow{5}{*}{\shortstack{Terephthalate \\ (H$_{4}$C$_{5}$O$_{2}$)}}
                              & NeuCBOT          &   0.247 & 0.226 & 0.082 & 0.216 & 0.130  \\
                              & SOURCES          &   0.172 & 0.159 & 0.041 & 0.151 & 0.075  \\
                              & G4-TENDL-2017    &   0.271 & 0.243 & 0.086 & 0.232 & 0.137  \\
                              & G4-JENDL         &   0.172 & 0.152 & 0.037 & 0.144 & 0.070  \\
                              & G4-JENDL$^{*}$   &   0.071 & 0.070 & 0.031 & 0.066 & 0.053  \\
\hline
\multirow{5}{*}{\shortstack{Viton \\ (H$_{2}$C$_{5}$F$_{8}$)}}
                              & NeuCBOT          &   19.2 & 17.0 & 3.53 & 16.1 & 8.53  \\
                              & SOURCES          &   13.5 & 12.4 & 2.17 & 11.6 & 5.51  \\
                              & G4-TENDL-2017    &   15.0 & 13.2 & 2.36 & 12.4 & 6.12  \\
                              & G4-JENDL         &   14.1 & 12.3 & 2.16 & 11.5 & 5.47  \\
                              & G4-JENDL$^{*}$   &   3.29 & 3.14 &  0.90 & 2.95 & 1.92 \\
\hline

\end{tabular}
\caption{Neutron yields (neutrons per 10$^{6}$ $\alpha$-decays) for the spectra of different decay chains in secular equilibrium, calculated with NeuCBOT, SOURCES and Geant4. The Geant4 calculations are performed with two different libraries: JENDL/AN-2005 (G4-JENDL) and TENDL-2017 (G4-TENDL-2017). We also show the contribution to G4-JENDL of the reactions that produce neutrons without emitting $\gamma$-rays (G4-JENDL$^{*}$).}
\label{tab:Westerdale_Comp}
\end{table*}

Concerning the energy spectra, we compare in figure~\ref{fig:Comp_WesterdaleSpectra} results from NeuCBOT and Geant4, using JENDL/AN-2005 and TENDL-2017. They correspond to the calculations performed with the $^{232}$Th decay chain in secular equilibrium with the materials in Table~\ref{tab:Westerdale_Comp}. The results of NeuCBOT and TENDL-2017, both from the TALYS code, are quite similar. JENDL/AN-2005 also produce similar spectra when the neutron production is dominated by F, Al or Si. On the contrary, the spectra obtained in Cirlex and Terephthalate are of a significantly lower energy than those obtained with NeuCBOT or TENDL-2017. This is because, as we pointed out in Section~\ref{Sec:dataLibs}, JENDL/AN-2005  underestimates the energy of secondary neutrons in some cases. Other comparisons between spectra obtained with the different codes are in~\cite{NeuCBOT,Comp_COOLEY,Comp_Fernandes,Comp_Selvi}.

\FloatBarrier

\section{Conclusions}

We have proved that Geant4 can be used to calculate neutron yields and energy spectra from ($\alpha$,n) reactions. Geant4 uses the information compiled in data libraries originally written in the ENDF-6 format, so that the quality of the results will be determined by the quality of the data in these libraries. One of the main advantages of Geant4 over other codes (NeuCBOT, SOURCES, NEDIS and USD) that perform similar calculations is that, if the information in the data library allows it, Geant4 can generate $\gamma$-rays in coincidence with  neutrons. Other advantages are the possibility of implementing geometries almost arbitrarily complex and that it allows to integrate in the same simulation the neutron generation and the neutron transport.

We have also reviewed the existing $\alpha$ particle incident ENDF-6 format databases, showing the differences and similarities among them. At this moment there are basically two different types: JENDL/AN-2005 and different versions of TENDL. JENDL/AN-2005 is an evaluated library and their ($\alpha$,n) cross sections rely on experimental data. It only contain information of 17 isotopes, but they are the most relevant ones for many practical situations. We also found that the energy of secondary neutrons is underestimated in some cases. The TENDL data libraries have been created from the output of the TALYS code, so they cover a much larger amount of isotopes ($\sim$2800), but results obtained with these libraries should be taken with care since they could deviate significantly from reality. For the isotopes tested, the cross sections in the different versions of the TENDL libraries considered (TENDL-2014, TENDL-2015 and TENDL-2017) do not differ more than 20\%. Each of them have two different sub-versions, one with all the non-elastic channels grouped together into a single channel in the entire energy range, and the other with explicit cross sections up to 30 MeV. This second sub-version allows to generate neutrons in coincidence with $\gamma$-rays, whereas the first one does not. Unfortunately, we have found that the TENDL sub-versions which include explicit reaction channels at low energies have not been constructed correctly for some isotopes.

Several neutron yields and energy spectra from ($\alpha$,n) reactions have been obtained with Geant4 and compared with experimental data and other codes. The neutron yields obtained with JENDL/AN-2005 agree better than 10\% with the experimental data in most of the tested cases. Yields from TENDL-2017 tend to be larger than the experimental ones, and the size of the difference varies from case to case. Concerning the other tested codes, the results from NEDIS-2.0 and SOURCES are similar to those from JENDL/AN-2005, the USD code underestimates systematically the yields considered, and NeuCBOT overestimates them by 60\% on average and up to more than a factor of 2 in some cases. The energy spectra from NeuCBOT and TENDL-2017 gave similar results, whereas JENDL/AN-2005 spectra have too low energies in some cases.

For some low-background experiments, it is important to distinguish between neutrons emitted in coincidence with $\gamma$-rays and neutrons generated without $\gamma$-ray emission. It is the latter that limit the sensitivity of detectors for dark matter direct search experiments. For this reason, we have calculated the contribution to the neutron yields of the reactions that produce neutrons without emitting $\gamma$-rays for some materials relevant for the mentioned experiments, using JENDL/AN-2005. For some cases, the obtained values are significantly lower than the (total) neutron yields obtained with other codes. For example, in the case of PTFE and Viton, these values are up to 6 and 4 times lower than the neutron yields from NeuCBOT and SOURCES, respectively.

\acknowledgments

This work was supported in part by the Spanish national company for radioactive waste management ENRESA, through the CIEMAT-ENRESA agreements on ``Transmutaci\'{o}n de radionucleidos de vida larga como soporte a la gesti\'{o}n de residuos radioactivos de alta actividad''; by the Spanish Plan Nacional de I+D+i de F\'{i}sica de Part\'{i}culas, through the projects FPA2016-76765-P and FPA2017-82647-P; and by the ``Unidad de Excelencia Mar\'{i}a de Maeztu: CIEMAT - F\'{i}sica de Part\'{i}culas'' through the grant MDM-2015-0509.


\bibliographystyle{JHEP}
\bibliography{EMendoza_G4AN_v01}

\end{document}